\documentclass{article}
\usepackage[T1]{fontenc}
\usepackage[latin1]{inputenc}
\usepackage{graphics}
\usepackage{graphicx}
\usepackage{longtable}
\usepackage{color}
\usepackage{pslatex}
\usepackage{hyperref}
\usepackage{sverb}
\usepackage{subfigure}
\usepackage{epstopdf}

\usepackage{multirow}
\usepackage{adjustbox}
\usepackage{indentfirst}

\usepackage{a4p}
\usepackage{rotating}

\usepackage{graphicx}
\usepackage{atlasphysics}
\usepackage{subfigure}
\usepackage{longtable}
\usepackage{multirow}
\usepackage{textcomp}
\usepackage{authblk,colortbl}
\usepackage{units}
\usepackage{amsmath,amssymb,amsfonts} 

\usepackage{authblk}

\usepackage{rotating}

\begin{document}
\begin{titlepage}
 
\begin{flushright} 
{ \bf IFJPAN-IV-2018-20 
} 
\end{flushright}
 
\vskip 30 mm

\begin{center}
  {\bf\huge  Machine learning classification: case of }\\
  \vskip 4mm
  {\bf\huge  Higgs boson CP state in $H \to \tau \tau$ decay at LHC}\\
\end{center}
\vskip 13 mm

\begin{center}
   {\bf K. Lasocha$^{a,b}$, E. Richter-Was$^{a}$, D. Tracz$^{a,*}$, Z. Was$^{c}$ and P. Winkowska$^{d,*}$}\\
   \vskip 3 mm
   {\em $^a$ Institute of Physics, Jagellonian University, ul. Lojasiewicza 11, 30-348 Krak\'ow, Poland} \\
   {\em $^b$ CERN, 1211 Geneva 23, Switzerland } \\
   {\em $^c$  IFJ-PAN, 31-342, ul. Radzikowskiego 152, Krak\'ow, Poland}\\
   {\em $^d$ Department of Computer Science, AGH USiT, Al. Mickiewicza 30, 30-059 Krak\'ow, Poland}
\end{center}
\vspace{1.1 cm}
\begin{center}
  {\bf   ABSTRACT  }
\end{center}

  Machine Learning (ML) techniques are rapidly finding a place among the
  methods of High Energy Physics data analysis. Different approaches are
  explored concerning how much effort should be put into building high-level
  variables based on physics insight into the problem, and when it is enough
  to rely on low-level ones,  allowing ML methods to find patterns without
  explicit physics model.

In this paper we continue the discussion of previous publications  on the CP state of
the Higgs boson measurement of the $H \to \tau \tau$ decay channel with the
consecutive 
$\tau^\pm \to \rho^\pm \nu$; $\rho^\pm \to \pi^\pm \pi^0$  and $\tau^\pm \to a_1^\pm \nu$;
$a_1^\pm \to \rho^0 \pi^\pm \to 3 \pi^\pm$ cascade decays.
The discrimination of the Higgs boson CP state is studied as a binary classification
problem between CP-even (scalar) and CP-odd (pseudoscalar), using {\it Deep Neural Network (DNN)}.
Improvements on the classification from the
constraints on directly non-measurable outgoing neutrinos are discussed.  
We find, that once  added, they  
enhance the sensitivity sizably, even if only imperfect
information is provided.
In addition to {\it DNN} we also evaluate and compare 
other ML methods: {\it Boosted Trees (BT)}, {\it Random Forest (RF)} and {\it Support Vector Machine (SVN)}.

\vskip 1 cm


\vfill
{\small
\begin{flushleft}
{   IFJPAN-IV-2018-20 
December 2018}
\end{flushleft}
}
 
\vspace*{1mm}
\footnoterule
\noindent
    {\footnotesize 

      $^{(*)}$  In time of working on this project.    

This project was supported in part from funds of Polish National Science
Centre under decisions DEC-2017/27/B/ST2/01391.
DT and PW were supported from funds of Polish National Science Centre under
decisions DEC-2014/15/B/ST2/00049.

Majority of the numerical calculations were performed at the PLGrid Infrastructure of 
the Academic Computer Centre CYFRONET AGH in Krakow, Poland.
    }
\end{titlepage}

\section{Introduction}
\label{Sec:Intro}

Machine Learning (ML) techniques find increasing number of applications in  High Energy Physics phenomenology.
With Tevatron and the LHC experiments, it became an analysis standard. The ML techniques
are used for event selection, event classification, background suppression for the signal events of the interest, etc.
For a comprehensive recent review see~\cite{Guest:2018yhq,Carleo:2019ptp,Albertsson:2018maf}. Over the last years
the most significant progress in phenomenology due to ML techniques, in particular recent development in neural network methods, was  in
hadronic jets reconstruction and classification: jet substructure, jet-flavour, jet-charge
 and jet-mass. They addressed successfully long-standing challenges of more classical algorithms,
see e.g. Refs.~\cite{Baldi:2016fql,Shimmin:2017mfk,Guest:2016iqz,Louppe:2017ipp,Fraser:2018ieu,Kasieczka:2019dbj,Kasieczka:2018lwf}.

In this paper we present studies on the seemingly related problem: exploration
how the substructure and pattern of hadronically
decaying $\tau$ leptons can be useful to determine the CP state of the
Higgs boson in the decay $H \to \tau \tau$. Theoretical description of the
process incuding $\tau$ lepton decays is relatively simple and of
the minor theoretical ambiguities only. On the other hand,
complicated detection approach remains a challenge. For example, indirect
constraints had to be devised and validated instead of non-measurable
$\tau$-neutrino momenta. Related part of the sensitivity was
often compromised. 

This problem has a long history~\cite{Kramer:1993jn,Bower:2002zx}. It was studied both for 
electron-positron~\cite{Rouge:2005iy,Desch:2003mw} and for hadron-hadron~\cite{Berge:2008dr,Berge:2015nua} colliders.
Despite some interest, CP in $H \to \tau \tau$ was not measured or
even explored in LHC  analysis designs.
While more classical experimental analysis strategies have been prepared and documented, see e.g.~\cite{ATL-PUB-2019-008}, for HL-LHC
strategies exploring the ML methods are still at the early stage. 

A typical experimental data sample consists of events. Each event can be understood as a point in a multi-dimensional coordinate space,
representing four-momenta and flavours of observed particles or groups of particles.
The physics goal is to identify properties of distributions
constructed from these events and to interpret them in physically meaningful way.
The 
ML algorithms with only the 
low-level features of the event 
are not necessarily able to capture efficiently all information available. The
best performing strategy still
seems to be 
 mixing of low-level information with the human-derived high-level features, based on the physics insight into the problem.
 Examples of such analyses are presented in~\cite{Baldi:2014kfa,Baldi:2014pta}.
 The strategy of mixing low-level and high-level features, prepared to remove trivial (physics-wise) symmetries
are explored successfully there. Then, the ML algorithms do not need to learn some basic physics rules, like rotation symmetry.

In the previous papers~\cite{Jozefowicz:2016kvz,Barberio:2017ngd} we have demonstrated that ML methods, like {\it Deep  Neural
Network}  {\it(DNN)}~\cite{deeplearningbook}, can serve as a promising analysis method to constrain the Higgs boson CP state in the  decay channel $H \to \tau \tau$.
We considered two decay modes of the $\tau$ leptons:
$\tau^\pm \to \rho^\pm \nu$, $\tau^\pm \to a_1^\pm \nu$, followed by $\rho^\pm \to \pi^\pm \pi^0 $ and $a_1^\pm \to \rho^0 \pi^\pm \to 3 \ \pi^\pm$.
This forms three possible hadronic final state configurations: $\rho^\pm \rho^\mp$,  $a_1^\pm \rho^\mp$
and $a_1^\pm a_1^\mp$, each accompanied by the $\tau$-neutrino pair.
The information about the Higgs boson CP state is encoded in the angles between outgoing decay products and angles between intermediate
resonances decay planes. From the early studies~\cite{Bower:2002zx,Przedzinski:2014pla} performed with the rather  classical
{\it optimal variable}~\footnote{For definition see e.g. \cite{Davier:1992nw}.}
approach, we have observed that
the best discrimination was achievable from features constructed in the rest-frame of the primary intermediate resonance pair of the $\tau$ decays,
with the $z$-axis
aligned to these resonances direction.
This idea was explored also in~\cite{Jozefowicz:2016kvz} and will be followed in this paper. We have investigated inputs consisting of mixed low-level
and high-level features. Many of the high-level features turned out to be
not necessary, but nevertheless provided benchmark results.
On the other hand, (post-fact seemingly simple) non-trivial choices for the representation of some low-level features was necessary to achieve any
significant result.

The studies presented in paper~\cite{Jozefowicz:2016kvz} were limited to
input from  the hadronic decay products, $\pi^\pm, \pi^0$;
no detector effects were taken into account.  That study was followed 
by a more systematic evaluation within the context of experimental analysis~\cite{Barberio:2017ngd}, namely applying simplified detector effects to
the input features.
The conclusions of~\cite{Jozefowicz:2016kvz} on the {\it DNN} method performance survived, and we will not
follow this evaluation in
the scope of the paper again.


Studies presented in~\cite{Jozefowicz:2016kvz} have shown, that
the case of  $\rho^\pm \rho^\mp$ followed by  $a_1^\pm \rho^\mp$
is the most sensitive to Higgs CP channel, and somewhat weaker sensitivity is achieved in
$a_1^\pm a_1^\mp$ case. Should all the decay channels be equally sensitive to the  Higgs CP state?
In~\cite{Kuhn:1995nn} it was demonstrated that yes, the sensitivity of 
each $\tau$ decay channel  to  spin is the same.
Unfortunately,
this requires control of all $\tau$ decay products momenta, in particular 
of non-measurable neutrinos. Studies presented in~\cite{Jozefowicz:2016kvz} did not rely on the
complete information,
limiting input information to the hadronic (visible) decay products only.
 However,
it is possible to overcome this limitation and reconstruct, with approximation, the neutrino
momenta from the  $\tau$ decay vertex position and event kinematics (momenta
of visible $\tau$ decay products, overall missing $p_T$ and overall
collision center of mass system energy).
 Such reconstruction is challenging from the experimental perspective and also
 for the analysis design: relations between necessary features are
 more complicated.
 Nevertheless this brings new opportunities for ML methods which we will explore with the
 help of {\it expert variables}: the azimuthal angles of the neutrino orientation.
The encouragement that the angle  may become  experimentally 
 available with adequate precision can be concluded from a recent experimental publication 
 of the LHC collaborations on the $H \to \tau\tau$ signal measurement~\cite{Aaboud:2018pen, Chatrchyan:2014nva},
 $\tau$ substructure reconstruction and classification~\cite{Aad:2015unr, Sirunyan:2018pgf}  and also progress
 on the precision of B meson decay vertex position measurements~\footnote{Expected performance of the $\tau$ decay vertices
 measurements based on the collisions data have not been published by LHC experiments yet.}~\cite{Aad:2015ydr, Sirunyan:2017ezt}.

 We attempt to reconstruct the two  neutrinos four-momenta
 (i.e. 6 quantities) from the experimentally available quantities
 and examine when such approximate information can be useful. 
To achieve this goal the following 3 steps are proposed:
\begin{enumerate}
\item
  reconstruction of neutrino 4-momenta components collinear to directions
  of visible decay products of $\tau$ leptons, from  the whole event
  missing transverse energy $E^x_{miss}, E^y_{miss} $
  and from the invariant mass of the Higgs boson $m_H$,
\item
  reconstruction of the transverse part of  neutrino momenta from  the $\tau$ lepton invariant mass
  constraint,
\item
  reconstruction of the two remaining azimuthal angles $\phi_{\nu_1}$, $\phi_{\nu_2}$
  of the neutrinos (or equivalent information); with the help of $\tau$-decay
   position vertices.
\end{enumerate}

After step (1) we have
4 independent variables to constrain, after step (2) only two remain.
The load on constraints
from the $\tau$ decay vertex position, probably the least precise to measure, is
minimized. This approach can be understood as an attempt to construct
high-level feature with the expert-supported design. This, if useful, may be later
replaced with better choices.
Several papers with optimal variables in mind followed such strategy \cite{Rouge:2005iy,Desch:2003mw,Berge:2015nua}.

For compatibility,
we use the same simulated samples as for Ref.~\cite{Jozefowicz:2016kvz}, namely Monte Carlo events of the Standard Model,
125~GeV Higgs boson, produced in pp collision at 13 TeV centre-of-mass energy, generated with {\tt Pythia 8.2}~\cite{Sjostrand:2014zea} and with
spin correlations simulated using {\tt TauSpinner}~\cite{Przedzinski:2014pla}.
For $\tau$ lepton decays we use {\tt Tauolapp}~\cite{Davidson:2010rw}. All spin and
parity effects are implemented with the help of {\tt TauSpinner}  weight $wt$. 
That is why the samples prepared for the CP even or odd Higgs are correlated.
For each  channel we use about $10^7$ simulated Higgs events~\footnote{This is 10 times larger statistics than used
in~\cite{Jozefowicz:2016kvz}.}. To partly emulate detector conditions, 
a minimal set of cuts is used. We require that the transverse momenta of the visible decay products combined,
for each $\tau$, are larger than 20 GeV. It is also required that the transverse momentum of each $\pi^\pm$ is larger than 1 GeV.

As in~\cite{Jozefowicz:2016kvz} we perform  {\it DNN} analysis for the three channels
of the
Higgs i.e. $\tau$ lepton-pair decays, 
denoted respectively as: $\rho^\pm-\rho^\mp$, $a_1^\pm-\rho^\mp$ and $a_1^\pm-a_1^\mp$.
%
Only two hypotheses on Higgs parity are confronted.
However, extension to parametrised classification,
similar to the approach taken in ~\cite{Baldi:2016fzo}, could be envisaged as an obvious next step;  the measurement of the Higgs CP parity  mixing angle.
Our paper can be also understood as a  work in that direction.

Our baseline for  ML methods is the {\it DNN}, nonetheless  we have also worked with more
classical  ML techniques
like {\it Boosted Trees} {\it (BT)}~\cite{Chen:2016btl}, {\it Random Forest} {\it (RF)}~\cite{Breiman:2001hzm} and
{\it Support Vector Machine} {\it (SVM)}~\cite{Bevan:2017stx}. 
The comparative analysis is presented for the $\rho^{\pm}-\rho^{\mp}$ case and for smaller events samples of about $10^6$ events.

Our paper  is organized as follows.
In Section~\ref{Sec:PhysRevD} we briefly recall the physics nature of the
problem and results from previous studies
of Ref.~\cite{Jozefowicz:2016kvz}.
In Section~\ref{Sec:addNeutrina} we discuss how to reconstruct, with some approximation,  the outgoing neutrinos momenta.
We exploit: collinear approximation, mass constraints, and information on spatial position
of production and decay vertices. 
In Section~\ref{Sec:DNNresults} we present an improvement in the {\it DNN} classification from information on the neutrinos.
We quantify what is the necessary precision on neutrinos azimuthal angle
to  improve the performance of the classifier.
In Section~\ref{Sec:Summary} the main results are sumarised and outlook is provided.

In Appendix~\ref{App:DNN} details concerning the {\it DNN} analysis implementation are given.
In Appendix~\ref{App:MLdivers} results achieved on the problem, but with
other ML techniques:
-- {\it BDT, RF} and {\it SVM} -- are presented.
We also discuss  technical performance, like usage of CPU or transient memory. 

\section{Classification based on hadronic decay products}
\label{Sec:PhysRevD}

Let us comment briefly on a few selected results~\footnote{In the present paper we were able to
improve with respect to~\cite{Jozefowicz:2016kvz} mainly thanks to 10 times larger training samples.}
from paper~\cite{Jozefowicz:2016kvz},
summarized in Table~\ref{Tab:PhysRevD}.
For the {\it DNN} classification, only directly measurable 4-momenta of the hadronic decay products of the $\tau$ leptons were considered.
They were boosted
to the rest-frame of the primary intermediate resonance pairs; respectively $\rho^\pm-\rho^\mp$, $a_1^\pm-\rho^\mp$ or $a_1^\pm-a_1^\mp$.
All four vectors were later rotated to the frame where primary resonances  were placed  along the $z$-axis.
It greatly improved the learning process. The {\it DNN} algorithm did not have
to e.g. rediscover rotational symmetry and
from the very beginning internal weights of the {\it DNN} algorithms could
determine transverse CP sensitive degrees of
freedom from the longitudinal ones. 
To quantify the performance for Higgs CP classification we have used
a weighted Area Under Curve (AUC) and Receiver Operator Characteristic (ROC) curve~\cite{roc-1,Fawcett2006}. For each simulated
event we know, from the calculated matrix elements,
the probability that an event is sampled as scalar or pseudoscalar
(for details see Appendix~\ref{App:DNN}).
This forms so called {\it oracle predictions},
i.e. ultimate discrimination for the problem which is about 0.782,
independently~\footnote{Consequence of $\tau$ decay dynamic. See e.g. Ref.~\cite{Kuhn:1995nn}.}
of the $\tau$ decay channels. Random classification corresponds to  0.500.

For  the studied
$\tau$-pair decay channels, the  AUC  in the   0.557 - 0.638
range was  achieved. 
Note, that so much lower than oracle predictions AUC  is due to
missing information on the neutrino momenta,
which are important carriers of the spin information, but 
are not accessible directly from the  measurement.
Let us explain very briefly the physics context of the problem.

The Higgs  boson Yukawa coupling expressed with the help of the
scalar--pseudoscalar parity mixing
angle $\phi$ reads as
\begin{equation}
  {\cal L}_Y= N\;\bar{\tau}{\mathrm  h}(\cos\phi+i\sin\phi\gamma_{5})\tau
\end{equation}
where
$N$ denotes normalization, $\mathrm h$ Higgs field and $\bar\tau$, $\tau$ spinors of the $\tau^+$ and $\tau^-$.  
The matrix element squared for the scalar / pseudoscalar / mix parity Higgs, with  decay
into $\tau^+ \tau^-$ pairs can be expressed as
\begin{equation} \label{eq:matrix}
  |M|^2\sim 1 +  h_{+}^{i} \;  h_{-}^{j} \; R_{i,j}; \;\;\;\;\; i,j=\{x,y,z\}
\end{equation}
 where  $h_{\pm}$ denote polarimetric
vectors of $\tau$ decays (solely defined by $\tau$ decay matrix elements) and
$R_{i,j}$ the density matrix of the $\tau$ lepton pair spin state.
In Ref.~\cite{Desch:2003rw} details of the frames used for the definition of $R_{i,j}$ and $h_{\pm}$
 are given.
 The corresponding CP sensitive spin weight $wt$ is simple:
\begin{equation} 
wt = 1-h_{{+}}^{z} h_{{-}}^{z}+ h_{{+}}^{\perp} R(2\phi)~h_{{-}}^{\perp}.
\end{equation}
The formula is valid for $h_\pm$ defined in $\tau^\pm$ rest-frames,
$h^{z}$ stands for longitudinal and  $h^{\perp}$ for  transverse component of $h$.
$R(2\phi)$ denotes  the matrix of $2\phi$ angle rotation  around the  $z$ direction:
$R_{xx}= R_{yy}={\cos2\phi}$, $R_{xy}=-R_{yx}={\sin2\phi}$.
The $\tau^\pm$ decay polarimetric vectors $h_{+}^i$,  $h_{-}^j$, in the simplest case
of $\tau^{\pm} \to \pi^{\pm} \pi^0 \nu $ decay, read 
\begin{equation}
h^i_\pm =  {\cal N} \Bigl( 2(q\cdot p_{\nu})  q^i -q^2  p_{\nu}^i \Bigr), \;\;\;  
\end{equation}
where  $\tau$ decay products  $\pi^{\pm}$, $\pi^0$ and $\nu_{\tau}$   4-momenta are denoted respectively as
$ p_{\pi^{\pm}}$, $p_{\pi^0}, p_{\nu}$ 
and $q=p_{\pi^\pm} -p_{\pi^0}$.
For $h_{\pm}^i$ of $\tau^\pm \to \pi^\pm\pi^\pm\pi^\mp \nu$ the
formula is longer, because dependence on modeling of the decay appear too \cite{Barberio:2017ngd}.
Obviously, complete CP sensitivity can be extracted
only if $p_{\nu}$ is known.  Note that the spin weight $wt$ is a simple first order trigonometric polynomial in a (doubled) Higgs CP parity mixing angle.
 This observation  is valid for all $\tau$ decay channels.

\begin{table*}
 \vspace{2mm}
  \begin{center}
  \caption{
    The {\it DNN} performance taken from~\cite{Jozefowicz:2016kvz} for discrimination
    between scalar and pseudoscalar Higgs CP state. For {\it DNN}
    classification only hadronic decay products 4-momenta 
    are used.\label{Tab:PhysRevD}}
 \vspace{2mm}
    \begin{tabular}{|l|r|r|r|}
    \hline
    Line content  & Channel: $\rho^\pm-\rho^\mp$          &  Channel: $a_1^\pm-\rho^\mp$                      &   Channel: $a_1^\pm-a_1^\mp$ \\
                  &             $\rho^\pm \to \pi^\pm \pi^0$ &  $a_1^\pm \to \rho^0 \pi^\pm, \rho^0 \to \pi^+ \pi^-$ & $a_1^\pm \to \rho^0 \pi^\pm, \rho^0 \to \pi^+ \pi^-$\\
                  &                                         &                     $\rho^0 \to \pi^+ \pi^-$        &               \\
    \hline
    Fraction of $H\to \tau \tau$  &  6.5\% & 4.6\% &   0.8\% \\
    \hline
    Number of features        &  24   & 32    &   48 \\
    \hline
    Oracle predictions         & 0.782 & 0.782 & 0.782 \\
    {\it DNN} classification  (AUC)        & 0.638 & 0.590 & 0.557 \\
    \hline
    \end{tabular}
  \end{center}
\end{table*}

\section{Approximating components of neutrino momenta}
\label{Sec:addNeutrina}

Our conjecture is that some of the steps listed in the introduction and presented
below may in the future be
replaced or optimized with  the solutions present in  ML libraries.
The expert variables, in particular  $\phi_{\nu_1}$, $\phi_{\nu_2}$, will not
be needed. We need to  explain our construction in detail first. 

We start with approximate neutrino momenta in the  ultra-relativistic
(collinear) approximation.
We temporarily assume that neutrino momenta and visible $\tau$ products momenta
are collinear to each other. Later we  relax this oversimplification.
This gives a reasonable approximation for collinear components which are
the largest ones (not only in the laboratory frame but also in the Higgs rest-frame and the rest-frame of its visible decay products).

\subsection{Collinear approximation}

The basic kinematical constraint on 4-momenta of each $\tau^{\pm} \to had^{\pm}  \; \nu$ decay reads ( $had^\pm$ stands for the hadronic system
produced in decay, i.e. $\pi^{\pm}, \pi^0$, etc. combined):
\begin{equation}
  \label{eq:collinear}
  p_{\tau_{1}}  = \ \ p_{{had}_{1}} + p_{\nu_{1}}, \ \ \ \ \ \ \ p_{\tau_{2}} = \ \ p_{{had}_{2}} + p_{\nu_{2}} 
\end{equation}
where
$ p_{\tau_{1}}, p_{\tau_{2}}$ denote 4-momenta of decaying $\tau$ leptons;  $p_{{had}_{1}}, p_{{had}_{2}}$ denote
4-momenta of their hadronic (i.e. measurable) decay products combined and  $p_{\nu_{1}}, p_{\nu_{2}}$ denote 4-momenta of the decay  neutrinos.

We temporarily assume that the directions of the hadronic decay products and neutrino are parallel
to the direction of the decaying $\tau$ and 
\begin{equation}
  \label{eq:collinear1}
  \vec p_{had}  = \ \ x \cdot \vec p_{\tau}, \ \ \ \ \ \ \ \vec p_{\nu} = \ \ (1 - x) \cdot  \vec p_{\tau}, 
\end{equation}
where $x$ is of the (0,1) range, then  for the $\tau^+$ and $\tau^-$
we can write:
\begin{equation}
  \label{eq:collinear2}
  \vec p_{\nu}  = \ \ \frac{1-x}{x} \cdot \vec p_{had} = \ \ \alpha \cdot  \vec p_{had}.
\end{equation}
From Eq.~(\ref{eq:collinear2}) we obtain
\begin{equation}
  \label{eq:NNobserv0}
  |\vec p_{\nu_{1}}|  = \ \ \alpha_1 \cdot |\vec p_{{had}_{1}}|, \ \ \ \ \ \ \ |\vec p_{\nu_{2}}|  = \ \ \alpha_2 \cdot |\vec p_{{had}_{2}}|. 
\end{equation}
These relations hold in the laboratory frame and in the rest-frame of
the hadronic decay products as well, which is   a consequence of properties of Lorentz
transformations of  ultra-relativistic particles. That is why we can  calculate
$\alpha_1$, $\alpha_2$  in the laboratory frame but use them
in the rest-frame of the hadronic decay products combined. That frame  
seems to be optimal~\cite{Jozefowicz:2016kvz} for the  construction of expert  variables for ML classification.


\subsubsection{The $E^x_{miss}, E^y_{miss} $ constraints}

The laboratory frame event momentum imbalance in the plane transverse
to the beam direction, usually denoted as $E_{miss}^x, E_{miss}^y$,
 can be used to constrain neutrino momenta.
It can be  attributed to the sum of transverse components of the neutrino  momenta, but it also
accumulates all imperfections of the reconstruction of the other outgoing particles of that event.
Then, thanks to
relation~(\ref{eq:collinear2}):
\begin{eqnarray}
  \label{eq:EXYmiss}
 E_{miss}^{x} && = \ \ p_{\nu_{1}}^x +  p_{\nu_{2}}^x  =  \ \ \alpha_1  \cdot p_{{had}_{1}}^x + \alpha_2   \cdot p_{{had}_{2}}^x, \\
 E_{miss}^{y} && = \ \ p_{\nu_{1}}^y +  p_{\nu_{2}}^y  =  \ \ \alpha_1  \cdot p_{{had}_{1}}^y + \alpha_2   \cdot p_{{had}_{2}}^y , \nonumber  
\end{eqnarray}

\begin{equation}
  \alpha_1 =  \frac{E_{miss}^{x} - \alpha_2   \cdot p_{{had}_{2}}^x }{p_{{had}_{1}}^x}
  \label{eq:missX}
\end{equation}
or
\begin{equation}
  \alpha_1 =  \frac{E_{miss}^{y} - \alpha_2   \cdot p_{{had}_{2}}^y }{p_{{had}_{1}}^y}.
   \label{eq:missY}
\end{equation}
Finally solving for $\alpha_1$, $\alpha_2$ we obtain  expressions
\begin{eqnarray}
  \label{eq:EmissConstr}
  \alpha_2  &&= \frac{ E_{miss}^{y} \cdot p_{{had}_{1}}^x -  E_{miss}^{x} \cdot p_{{had}_{1}}^y} { p_{{had}_{2}}^y \cdot p_{{had}_{1}}^x - p_{{had}_{2}}^x \cdot p_{{had}_{1}}^y}, \\
  \alpha_1  &&= \frac{E_{miss}^{x} - \alpha_2  \cdot p_{{had}_{2}}^x }{p_{{had}_{1}}^x}, \nonumber
\end{eqnarray}
useful for the studies of ML classification.

\subsubsection{ Using $m_{H}$ constraint}

Equations (\ref{eq:EmissConstr}) alone provide solution for  $\alpha_1$
and $\alpha_2$. However, input
$E^x_{miss},E^y_{miss} $ have  large experimental uncertainties. At the same time,
the high quality  constraint from the known Higgs-boson and  $\tau$-lepton masses
is available
{\small
\begin{eqnarray}
 && m^2_H = ( p_{\tau_{1}} +  p_{\tau_{2}} )^2 = 2 \cdot m_{\tau}^2 + 2 \cdot (1 + \alpha_1) \cdot (1 + \alpha_2) \\
&&[E_{{had}_1}E_{{had}_2} - p_{{had}_1}^x \cdot p_{{had}_2}^x - p_{{had}_1}^y \cdot p_{{had}_2}^y -  p_{{had}_1}^z \cdot p_{{had}_2}^z].\nonumber 
\end{eqnarray}%
}%
   $E_{{had}_1}$, $E_{{had}_2}$ denote  the hadronic systems ${had}_1$ and ${had}_2$ energies.
Later we will use similar notation  $E_\nu$ 
for the neutrino energy.

Unfortunately,
 only the product $(1+\alpha_1) \cdot (1+\alpha_2)$ can be controlled in this way,
 {\small
\begin{eqnarray}
  &&  (1 + \alpha_1) \cdot (1 + \alpha_2) = \label{eq:MHiggs} \\
  &&\frac{m^2_H/2 - m_{\tau}^2}{E_{{had}_1}E_{{had}_2} - p_{{had}_1}^x \cdot p_{{had}_2}^x - p_{{had}_1}^y \cdot p_{{had}_2}^y -  p_{{had}_1}^z \cdot p_{{had}_2}^z}. \nonumber 
\end{eqnarray}
}

\subsubsection{Choosing optimal solution for longitudinal neutrino momentum}
\label{Sec:Optimal}
To constrain $\alpha_1,\alpha_2$
we have  three independent equations of (\ref{eq:EmissConstr}) and (\ref{eq:MHiggs}) at our disposal.
%
We have checked that all three options:

\begin{itemize}
\item
{\it Approx-1} : formulae (\ref{eq:EmissConstr}) only,
\item
{\it Approx-2} : formula (\ref{eq:MHiggs}) and $\alpha_1$ from formulae  (\ref{eq:EmissConstr}),
\item
{\it Approx-3} :  formula (\ref{eq:MHiggs}) and $\alpha_2$ from formulae   (\ref{eq:EmissConstr}),
\end{itemize}
lead to comparable predictions and marginal differences of the ML performance
at least as long as measurement ambiguities of
 $E_{miss}^{x}$,  $E_{miss}^{y}$ are not taken into account.
It will be of concern for experimental precision. For now,
 the option {\it Approx-1} is  chosen
 as a base-line for the results~%
\footnote{This point may become important for discussion of ambiguities due to missing $p_T$ of jets accompanying the Higgs
  production. Then, it may be helpful to have 3 constraints which may be used e.g.
  for missing $p_T^{miss}$ generated by jets heavy
  flavour resonances of decays with neutrinos, contribute to $E_T^{miss}$ as well.} without much elaboration.

\begin{figure}
 \begin{center}                              
 { 
   \includegraphics[width=7.5cm,angle=0]{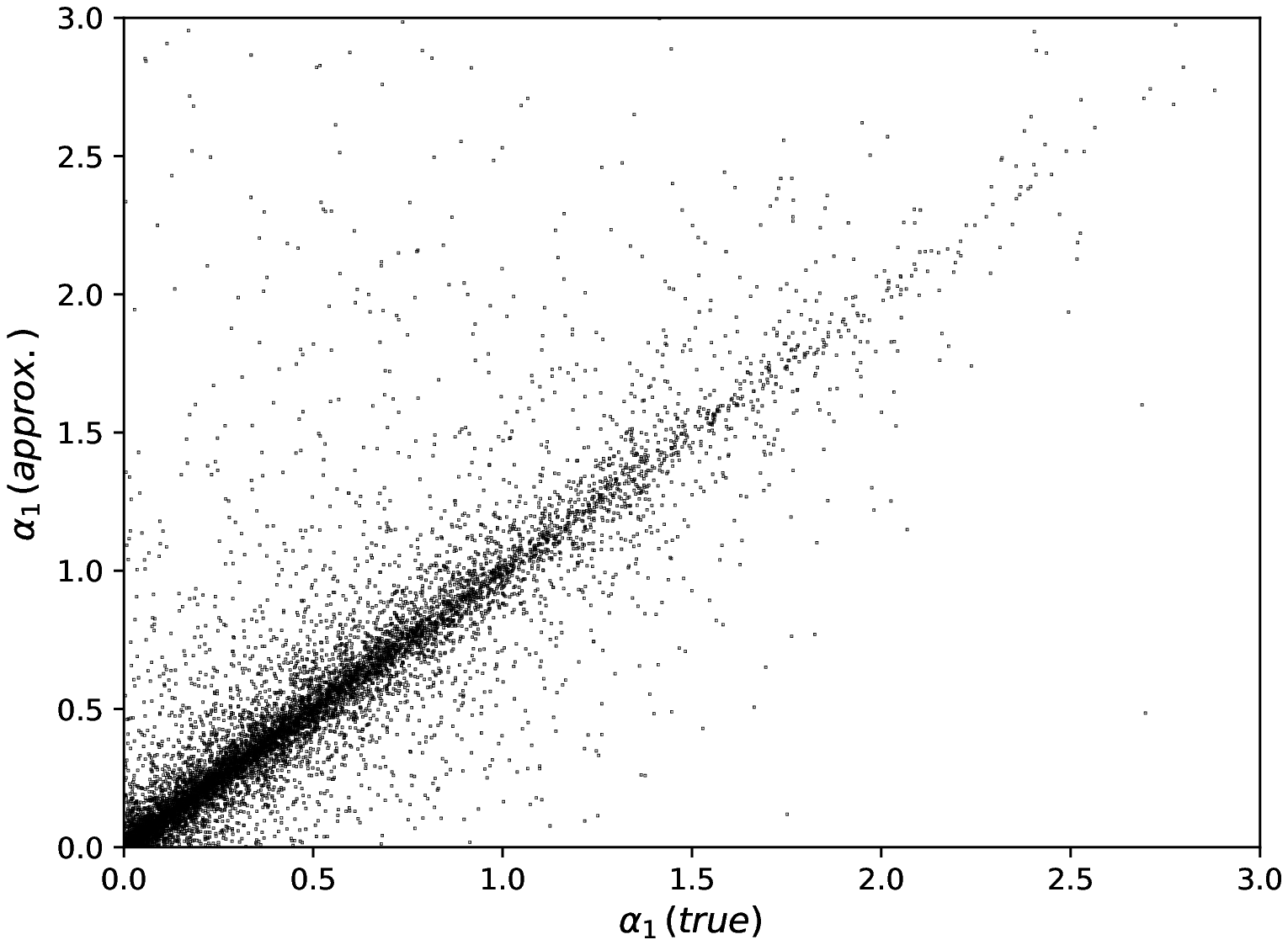}
   \includegraphics[width=7.5cm,angle=0]{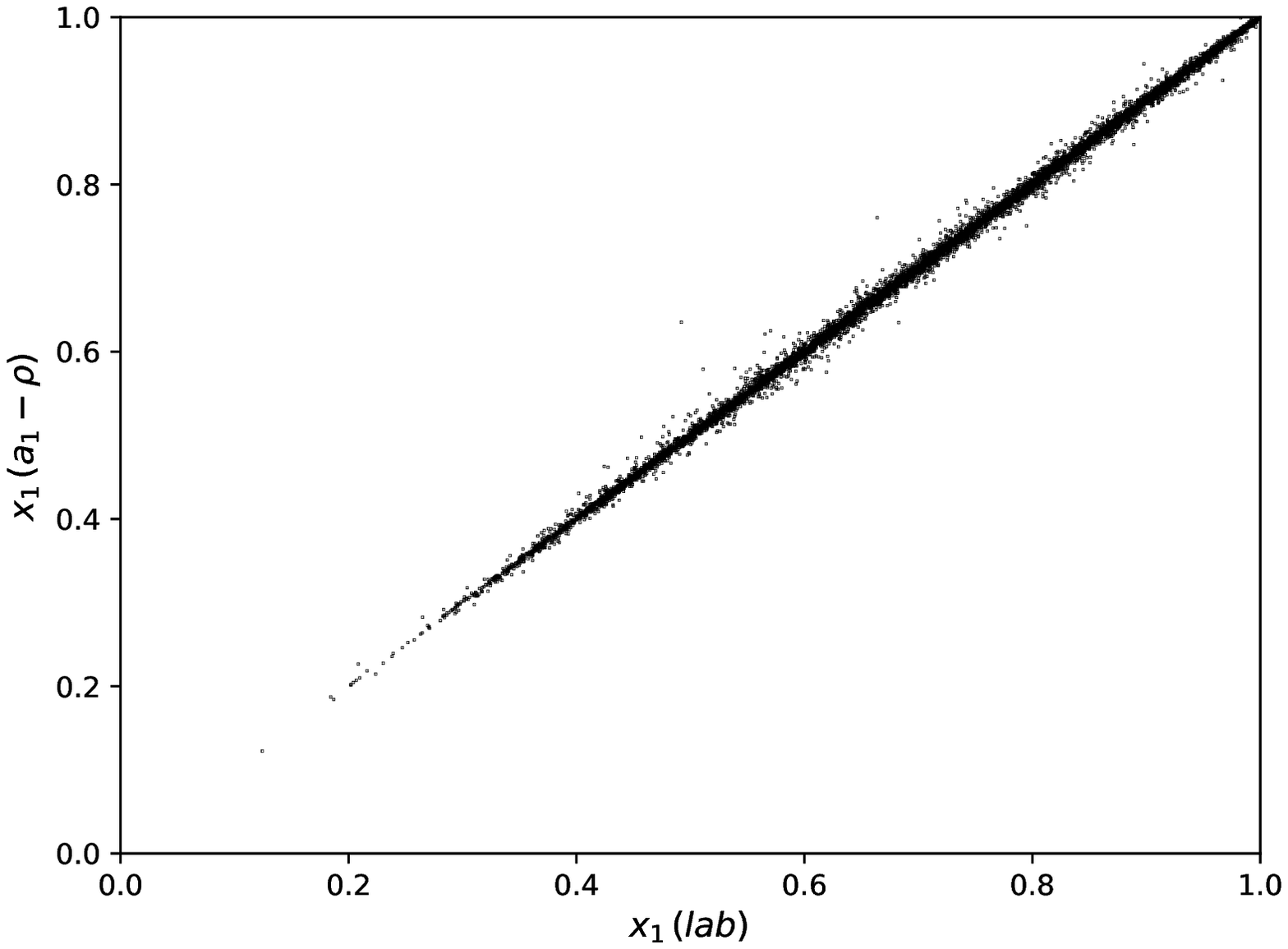}
}
\end{center}
 \caption{Left-side plot: correlation between true and approximated $\alpha_1$ calculated for the $a_1^\pm - \rho^\mp$.
  right-side plot: Correlation between fraction $x$ of $\tau$ lepton momentum carried by hadronic decay products obtained 
   in the {\it Approx-1} approximation; in the $a_1^\pm - \rho^\mp$ and laboratory frames.
   \label{fig:alpha1x1}}
\end{figure}

To illustrate the effectiveness,  the 
correlation between $\alpha_1$-true~\footnote{True or truth level is a short-cut denoting that it is calculated from the generated event kinematic, without any
approximation or smearing.}
and $\alpha_1$-{Approx-1} is shown
in Fig.~\ref{fig:alpha1x1} for the $a_1^\pm - \rho^\mp$ case (left plot).
In the right plot, as
consistency check, the correlation  of the $a_1^\pm - \rho^\mp$ rest-frame and  laboratory
frame energy fraction $x_1$, calculated  from $\alpha_1$ of
{\it Approx-1},  is given. A sample of $10^4$ events was used for these scattergrams.
The fraction of events contained in the band of $\Delta \alpha_1/\alpha_1 =  \pm 5\% (\pm 10\%$) is about 25\%(39\%)
and in the band  $\Delta x_1/x_1 = \pm 1\%$ is about 85\%. This relatively poor resolution in $\alpha_1$ will be
reflected in the resolution of approximated neutrino momenta. It will be interesting to observe how much
it will affect the classification capability of trained {\it DNN}, which will be discussed in Section~\ref{Sec:DNNresults}.  

\subsection{Energy and transverse component of neutrino momenta}

Now, with the help of approximated $p_{\nu}^z$ (the component longitudinal
along visible decay products),  we can turn our attention
to $p_{\nu}^x$ and $p_{\nu}^y$.
In the hadronic decay products system  rest-frame
the $p_{{had}_{1,2}}$ momenta are set along the $z$ direction thus $ p_{had}^x = p_{had}^y = 0 $.
The $\tau$ mass constraint reads 
\begin{equation}
m_\tau^2= (E_\nu+E_{had})^2-(p_{\nu}^x )^2-(p_{\nu}^y )^2-(p_{\nu}^z+p_{had}^z)^2,
\end{equation}
and for massless $\nu_\tau$  
\begin{equation}
0= (E_\nu)^2-|p^T_\nu|^2-(p^z_\nu)^2.
\end{equation}
The equations lead to the following relations:
\begin{eqnarray}
 \label{eq:EnuPT}
  E_{\nu}  && = \ \ \frac{m_{\tau}^2 - E_{had}^2 + (p_{had}^z)^2 + 2 \cdot p_{\nu}^z p_{had}^z}{2 E_{had}},\\
  p^T_{\nu} && = \ \ \sqrt{E_{\nu}^2 - (p_{\nu}^z)^2}, \nonumber
\end{eqnarray}
where for $p_{\nu}^z = \alpha \cdot p_{had}^z$  one of the  
$\alpha$ approximations from Section~\ref{Sec:Optimal}  is used.

\begin{figure} 
  \begin{center}                              
    { 
      \includegraphics[width=7.5cm , angle=0]{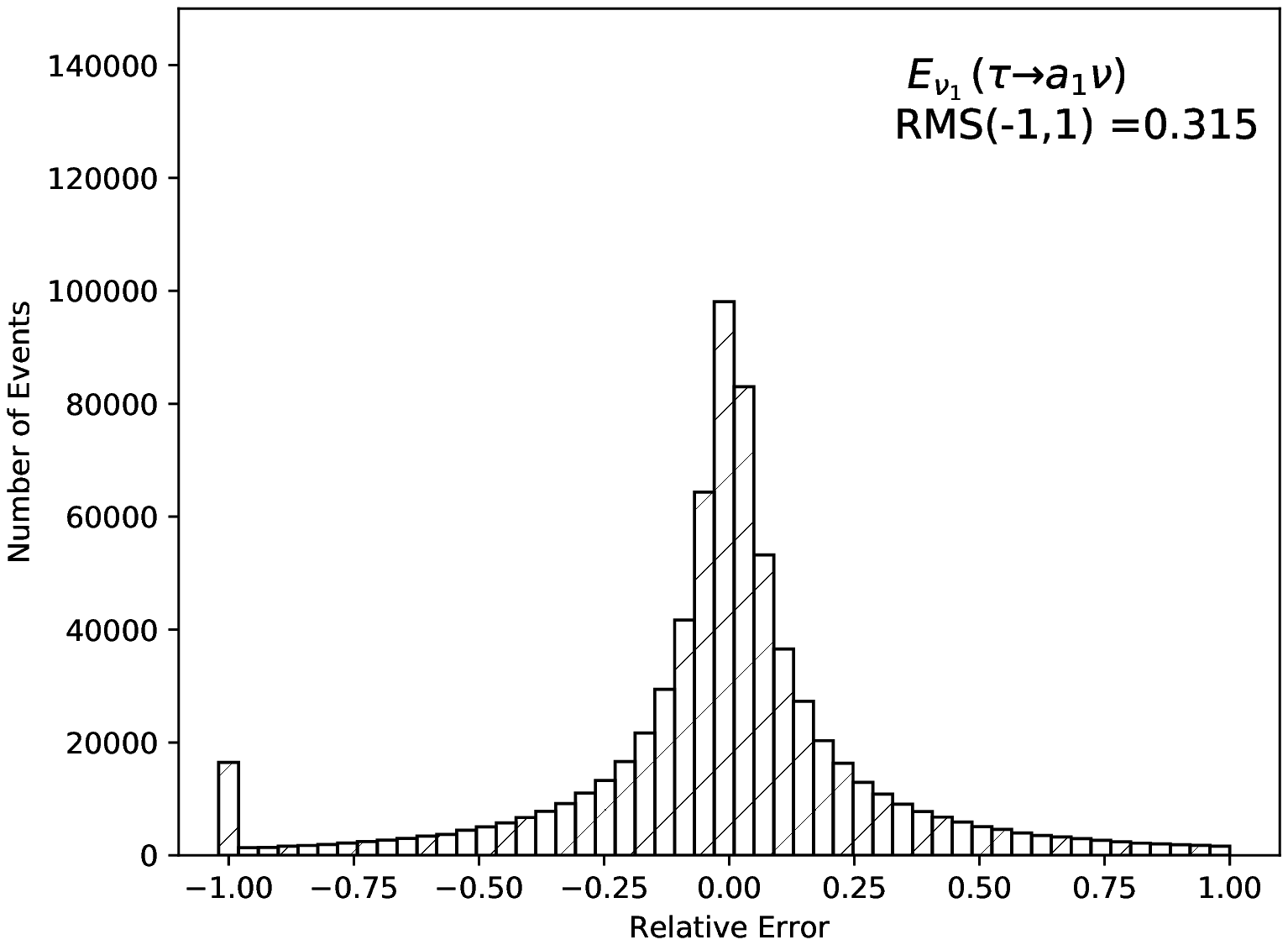}
      \includegraphics[width=7.5cm , angle=0]{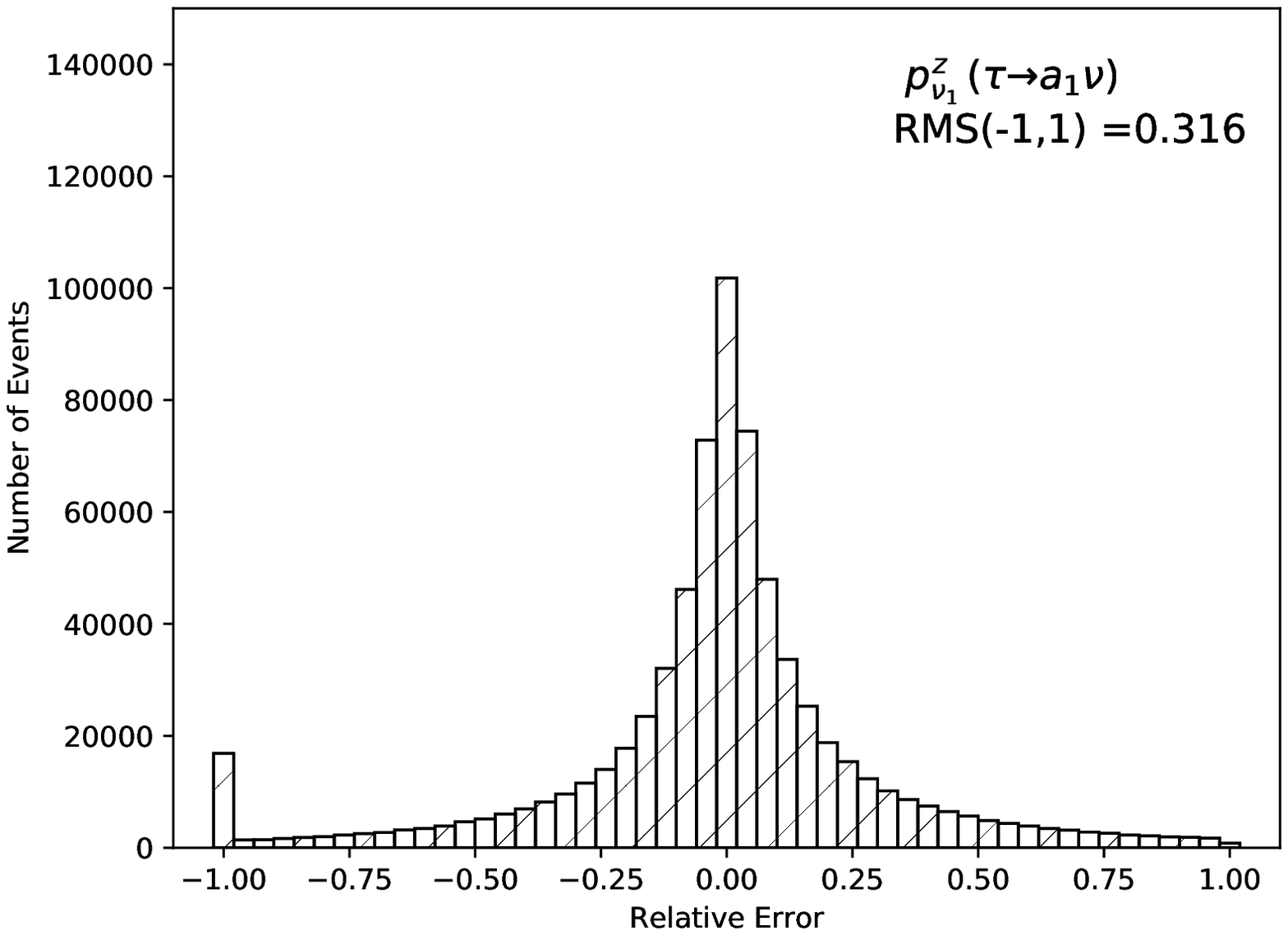}
      \includegraphics[width=7.5cm , angle=0]{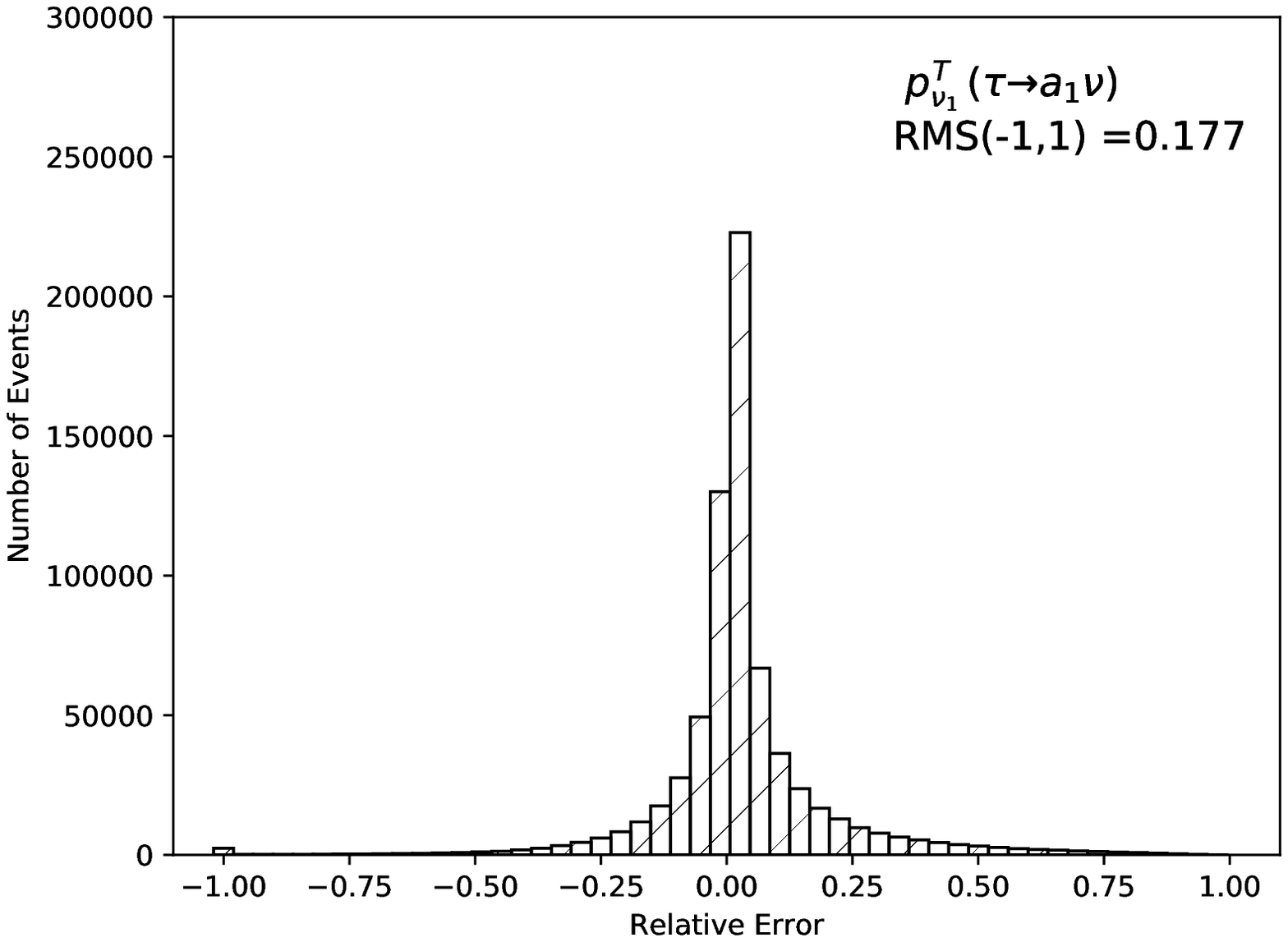}
    }
  \end{center}
  \caption  {
    The  $(x_{true} - x_{approx.})/x_{true}$ distribution for the
    reconstructed neutrino
    energy (left), longitudinal (right) and transverse (bottom)  momenta; 
    {\it Approx-1} was used. 
    Events of shifts  outside  (-1.0, 1.0) window are cumulated at -1.0 bin.
    The standard deviation denoted as (RMS)  is calculated for the (-1.0, 1.0) range.
    \label{Fig:Pnuresol}
  }
\end{figure}

The $\alpha_1, \alpha_2$,  $E_{\nu_1}$ and  $E_{\nu_2}$ must be positive,
otherwise the approximation fails and the event can not be used.
Also events with negative  approximated $(p^T_{\nu})^2$ could  be rejected,
but for our studies  we decided to set this component to zero
instead.
In total, about $17\%$ events are rejected for {\it Approx-1}. An
additional $11\%$ are rejected when for each event it is requested
that  with {\it Approx-2} and  {\it Approx-3}  the above criteria are
also fulfilled.
 In Fig.~\ref{Fig:Pnuresol}, the  distribution of relative shifts from generated to approximated $E_{\nu}, p^z_{\nu}, p^T_{\nu}$ is given for the 
$a_1^\pm - \rho^\mp$  case. The  $p^T_{\nu}$ is  approximated better 
than  $E_{\nu}, p^z_{\nu}$.
%
We remain encouraged because for ML classifications
even approximate observables
(expert variables)
may be useful to improve classification scores.

\subsection{Azimuthal angles of neutrinos}
\label{Sec:IP}

At this point, we are left without  two azimuthal angles for the orientation of
$p^T_{\nu_1}$ and $p^T_{\nu_2}$ only. To capture the sensitivity of the Higgs boson CP they have to
be known, preferably  in  the visible $\tau$-pair decay products rest-frame.
Those two angles can be inferred from 
the $\tau$ decay vertices positions and then through boosts and rotations
related to the azimuthal angles in visible decay products frame.

The transverse coordinates of the primary interaction point are to a good precision consistent with
zero. At the same time, the tracks of the $\tau$ decay products will not point
to this interaction vertex but to the  position of 
the $\tau$ decay vertex  shifted by the $\tau$ flight.
The direction of the $\tau$ flight can be  reconstructed, 
and as a consequence, so can be  its  momentum components. This provides 
a constraint on the $\nu_\tau$ momentum as well. We do not intend to go into details
of this challenging  secondary vertex position measurement.
Let us point to  Ref.~\cite{Aad:2015ydr,Sirunyan:2017ezt}, which discuss a
similar problem of
secondary vertex in case of B-meson decay and its application for
the  classification of hadronic 
jets. One may assume that such measurement is possible
for a $\tau$ lepton, and that the orientation of $\nu_\tau$ momentum around the
direction of visible hadronic $\tau$ decay products can be constrained.

To access how precisely we need to know this information we  take the
true azimuthal angles $\phi_{\nu_1}$, $\phi_{\nu_2}$
in the rest-frame of visible decay products and smear them.
For~~$ \Delta \phi_{\nu} = |\phi_{\nu}^{smear} - \phi_{\nu}^{true}| $
 smearing probability we take 
\begin{equation}
  f_{train}(\Delta \phi_{\nu}, \beta) = \frac{1}{\beta} \ exp(- \frac{1}{\beta} \Delta \phi_{\nu}).
  \label{eq:IP}
\end{equation}
 We have chosen the exponential shape,
 instead of often used Gaussian shape~\footnote{%
The Gaussian shape, as can be concluded  from central limit theorem,
the universal statistical
distribution  for the variable obtained as an average
of  large set  of independent stochastic variables, may be too
simplistic to use as a test example. Also, because the lifetime of the
$\tau$ follows an  exponential distribution we have chosen such a shape for the
smearing of $\phi_\nu^{smear}$ which can be, in real detector response
simulations, proportional to the (inverse) of the distance between
$\tau$ decay and production vertices. We have used other distributions
(like Gaussian distribution with exponentially enhanced tail) as well
and conclusions on  the {\it DNN} algorithm performances remained similar.
The {\it DNN} could learn from inaccurate distributions. The exponential tail
was supposed to introduce a penalty for learning. 
The actual
distribution depends on details of the detectors and reconstruction algorithms
used by the experiments. This challenging effort, requiring all details
of its geometry, has not been  completed so far.
Preliminary results are not available
publicly, in fact they are of considerable complexity and depend on the
detector regions (barrel, endcap etc.)  as well as on the  $\tau$ lepton energy
and decay channel.%
 }.
 Note however that the length
of the $\tau$ flight path follows an exponential distribution. 
We choose the  sign for the shift with equal probabilities.

We think, that at present it is premature to attempt realistic detector smearing.
Only in case of $\tau^\pm \to \pi^\pm \pi^\pm \pi^\mp \nu$ decay channel of the $Z/\gamma^* \to \tau\tau$
production, at LHC such attempts on investigating experimental smearings
for secondary vertex position are reported \cite{Cherepanov:2018npf}.
        
\subsection{Ansatz for  direction of the $\tau$ leptons}
\label{Sec:taudir}

In Subsection~\ref{Sec:IP} above we have discussed the possibility of adding
 approximate
 information on the angle of the outgoing neutrino in the decay plane
 to the features list.
However, for the  multi-variate methods, this angle does not have to be
present explicitly in
the feature list.
In fact, indirect information such as approximated
direction of the outgoing $\tau$ lepton may be good enough.

From the primary and secondary vertex positions,
direction of the laboratory system $\tau$ lepton momentum,
i.e. $p_x, p_y, p_z$, is constrained. Assuming  known 
$\tau$ time of flight ($t_{flight}$) and  mass $m_{\tau}$, we  calculate 
\begin{equation}
p_{i}^{\tau}= m_{\tau} \cdot (i_{sec.\ vtx}-i_{prim.\ vtx} )/t_{flight}, 
\end{equation}
where $i_{sec.\ vtx}, i_{prim.\ vtx}$ denote spatial position of the reconstructed
primary and secondary vertex respectively in the laboratory (collision) frame
($i=x,y,z$).
Instead of unknown true time-of-flight, we use the one of PDG~\cite{Tanabashi:2018oca}:
$c \tau_{\tau} = 87 \mu m$.
The true time-of-flight behaves according to the exponential distribution with 
mean  $<t_{flight}>$ = $\tau_{\tau}$. It imposes that the approximation 
used for estimating  $p_x, p_y, p_z$ is also characterised by an exponential distribution,
with mean and sigma close to their true values. The energy of the $\tau$ lepton
is then calculated using $\tau$ mass constraint
\begin{equation}
  E^{\tau} = \sqrt{(p_x^{\tau})^2 + (p_y^{\tau})^2 + (p_z^{\tau})^2 + m_{\tau}^2}.
\end{equation}

Now, the complete 4-momentum of each $\tau$  is
boosted into the $\rho^{\pm}-\rho^{\mp}$, $a_1^{\pm}-\rho^{\mp}$ or $a_1^{\pm}-a_1^{\mp}$
system rest-frame and added to the feature lists for {\it DNN} training.
{

\section{Classification with {\it DNN}}
\label{Sec:DNNresults}

The structure of the data and neural network architecture follows~\cite{Jozefowicz:2016kvz}. We start from the code used there.
For the convenience of the reader, we summarise the technical description of our {\it DNN} model in Appendix~\ref{App:DNN}.

Simulated data consist of events where all decay products are stored
together with their flavours. The four-momenta of the
 laboratory frame are stored and, whenever it is needed, transformed to respective rest-frames as explained in Section~\ref{Sec:PhysRevD}.
With respect to the analysis published in~\cite{Jozefowicz:2016kvz} we explore approximate information
on neutrino momenta derived from the kinematical constraints of the Higgs decay
products. We show that significant improvement may originate from even very
inaccurate information on the azimuthal angles of the neutrinos' 
directions.

We explore the potential of classification with the {\it DNN} technique with several variants of the feature lists
as detailed in Table~\ref{Tab:ML_variants}. They are grouped and marked as
{\tt Variant-X.Y}, where {\it X}  labels a
choice of the  main features and  {\it Y}  in most cases labels if they are calculated from the generator-level 4-momenta
or from the approximation; it may also mark if additional, high-level,
variables were used. It gives us very useful handles to quantify how much of the {\it DNN} performance we are loosing
due to certain approximations made on the groups of features.

\begin{table*}
 \vspace{2mm}
  \caption{
    Lists of features for ML classification, marked as {\tt Variant-X.Y}.
    In the third column, number  of features respectively for the
    $\rho^{\pm}-\rho^{\mp}$, $a_1^{\pm}-\rho^{\mp}$ and $a_1^{\pm}-a_1^{\mp}$ channels  are given.
    All components of the 4-momenta are taken in the hadronic decay
    products rest-frame. Primary resonances ($\rho^\pm$, $a_1^\pm$) aligned
    with the $z$ axis.
    The $E^x_{miss}$, $E^y_{miss}$ are of the laboratory frame.
    In practice, instead of $p_\nu^T$ and $\phi_\nu$, the
    pair of variables $p_\nu^T \cos\phi_\nu$, $p_\nu^T \sin\phi_\nu$ is used.
    \label{Tab:ML_variants}}
   \begin{center}
    \begin{tabular}{|l|l|c|l|}
    \hline
  Notation   & Features  & Counts & Comments\\
    \hline
 {\tt Variant-All}       &  4-momenta ($\pi^\pm, \pi^0, \nu$)                                                & 24/28/32 &  \\
    \hline
 {\tt Variant-1.0}       &  4-momenta ($\pi^\pm, \pi^0$)                                                     & 16/20/24 & as in Table 3 of \cite{Jozefowicz:2016kvz}\\
 {\tt Variant-1.1}       &  4-momenta ($\pi^\pm, \pi^0, \rho^\pm, a_1^\pm$), $m_i^2, m_k^2, y_i, y_k, \phi^*_{i,k}$  & 29/46/94 & \\
    \hline
 {\tt Variant-2.0}       &  4-momenta ($\pi^\pm, \pi^0$), $E_{\nu}$, $p^z_{\nu}$,$p^T_{\nu}$                    & 22/26/30 & \\
 {\tt Variant-2.1}       &  4-momenta ($\pi^\pm, \pi^0$), $E_{\nu}$, $p^z_{\nu}$,$p^T_{\nu}$                    & 22/26/30 & Approx. $E_{\nu}, p_{\nu}^z, p_{\nu}^T$\\
 {\tt Variant-2.2}       &  4-momenta ($\pi^\pm, \pi^0$), $E_{\nu}$, $p^z_{\nu}$,$p^T_{\nu}$, $E^x_{miss}$, $E^y_{miss}$  & 24/28/32 & Approx. $E_{\nu}^z, p_{\nu}, p_{\nu}^T$\\
    \hline
 {\tt Variant-3.0.0}    &  4-momenta ($\pi^\pm, \pi^0$), $E_{\nu}$, $p^z_{\nu}$, $p^T_{\nu}$,  $\phi_{\nu}$ & 24/28/32 & Approx. $E_{\nu}, \vec{p_{\nu}}$ \\
 {\tt Variant-3.1.$\beta$}    &  4-momenta ($\pi^\pm, \pi^0$), $E_{\nu}$, $p^z_{\nu}$, $p^T_{\nu}$,  $\phi_{\nu}$ & 24/28/32 & Approx. $E_{\nu}, \vec{p_{\nu}}$; $\phi_{\nu}$ smeared with  $\beta$ \\
 \hline
 {\tt Variant-4.0}             &  4-momenta ($\pi^\pm, \pi^0$, $\tau^\pm$)                                   & 24/28/32 & \\
 {\tt Variant-4.1}            &  4-momenta ($\pi^\pm, \pi^0$, $\tau^\pm$)                                   & 24/28/32 & Approx. $p_{\tau} $\\
 \hline
    \end{tabular}
  \end{center}
\end{table*}

In Table~\ref{Tab:ML_AUC}, we collect AUC scores and Average Precision Scores (APS)~\cite{Zhang2009},
obtained on the test sample of simulated data
(i.e. events not used for training or validation) with the {\it DNN}
trained on 50 epochs and with dropout = 0.20. Both are comparable, the APS score being systematically slightly lower,
except for few cases of the $a^{\pm}_{1}-a^{\mp}_{1}$ channel.
This configuration was found as most stable for comparison
of {\tt Variant-X.Y} classifications, but not necessarily represents the optimal performance of the
particular variant of the features list. 
In the first line of  Table~\ref{Tab:ML_AUC} we recall
the oracle predictions~\footnote{Because of physics properties,
  they should be the same for all channels, but as we are filtering events
  and finite statistics, they differ on third digit.}, for details see Appendix \ref{App:DNN}.
It cannot be outperformed by the {\it DNN} of  any {\tt Variant-X.Y}. It may not be reached even with a features list
containing the complete set of 4-momenta of $\tau$ decay products, denoted as {\tt Variant-All}.

In the following subsections we discuss those results in detail.

\begin{table}
 \vspace{2mm}
  \caption{
    The AUC and APS scores to discriminate  scalar from pseudo-scalar 
    CP state of the Higgs boson, obtained on the  test sample. The {\it DNN} was
    trained on 50 epochs and dropout=0.2 (except explicitly marked  case of {\tt Variant-All}).
    Results for $\rho^{\pm} - \rho^{\mp}$, $a_1^{\pm}-\rho^{\mp}$ and $a_1^{\pm}-a_1^{\mp}$
    channels are given.
    The first column labels choice of features.
     For details see Table~\ref{Tab:ML_variants}. 
    \label{Tab:ML_AUC}}
      \begin{center}
    \begin{tabular}{|l|c|c|c|}
    \hline
    Features  & AUC/APS  & AUC/APS   & AUC/APS   \\
    list &  ($\rho^{\pm}-\rho^{\mp}$) &  ($a_1^{\pm}-\rho^{\mp}$)  &  ($a_1^{\pm}-a_1^{\mp}$)  \\
    \hline
    Oracle predictions              & 0.784/0.785   &  0.781/0.783 &  0.780/0.782 \\
    \hline
 {\tt Variant-All}  (drop=0.0)         & 0.784/0.786  &   0.778/0.778 &  0.773/0.774 \\
 {\tt Variant-All}                        & 0.769/0.764  &   0.748/0.742 &  0.728/0.720  \\
    \hline
 {\tt Variant-1.0}                        & 0.655/0.654  &   0.603/0.602 & 0.573/0.578 \\
 {\tt Variant-1.1}                        & 0.656/0.655  &   0.609/0.607 & 0.580/0.585 \\
    \hline
 {\tt Variant-2.0}                        & 0.663/0.663  &   0.626/0.625 & 0.594/0.595  \\
 {\tt Variant-2.1}                        & 0.664/0.666  &   0.622/0.622 & 0.591/0.593  \\
 {\tt Variant-2.2}                        & 0.664/0.666 &   0.622/0.622 & 0.591/0.593 \\
    \hline
 {\tt Variant-3.0.0}                     & 0.771/0.771 &   0.749/0.743 & 0.728/0.721 \\
 {\tt Variant-3.1.2}                     & 0.760/0.759 &   0.738/0.730 & 0.718/0.710 \\
 {\tt Variant-3.1.4}                     & 0.738/0.735 &   0.714/0.705 & 0.687/0.677 \\
 {\tt Variant-3.1.6}                     & 0.715/0.713  &   0.689/0.680 & 0.660/0.652 \\
 \hline
 {\tt Variant-4.0}                        & 0.769/0.766  &   0.748/0.742 & 0.728/0.720 \\
 {\tt Variant-4.1}                        & 0.738/0.733 &   0.704/0.696 & 0.683/0.676  \\
    \hline
    \end{tabular}
  \end{center}
\end{table}	

\subsection{Benchmarks using all or only hadronic decay products}

For the first benchmark each event is represented with
4-momenta of both $\tau$-leptons decay products
(including neutrinos)  in the rest-frame of all hadronic decay products
combined. This set of features is denoted as {\tt Variant-All}.
Results are displayed in the second and third line of  Table~\ref{Tab:ML_AUC}.
The {\it DNN} should be able to reproduce oracle predictions, which is almost the case
if dropout is not used, but only approaches it with base-line configuration of dropout=0.20.
The dropout is lowering {\it DNN} performance in {\tt Variant-All}, but we have verified that for other feature lists
it is not always the case. It helps with suppressing
overfitting, as illustrated in Fig.~\ref{FigApp:DNN_dropout} of Appendix~\ref{App:DNN}.
In Fig.~\ref{Fig:ML_Variant-All-1.0}, left plot, we show for
the  $a_1^{\pm}-\rho^{\mp}$ channel {\tt Variant-All}, the 
AUC score as a function of number of epochs used for training and validating.
The scores up to about $0.75$ are reached for the validation sample and {\tt Variant-All}.

For the second benchmark following Ref.~\cite{Jozefowicz:2016kvz}, the same
events but with features limited to 4-momenta of visible $\tau$
leptons decay products and  quantities derived directly from them are used
~\footnote{Main results of Ref.~\cite{Jozefowicz:2016kvz},
  where only {\tt Variant-1.X} were
studied have been recalled in Section~\ref{Sec:PhysRevD}, Table~\ref{Tab:PhysRevD}.
Nonetheless for overall consistency, we have reevaluated some of those results again.
}.
The set with only 4-momenta of visible decay products in the respective rest-frames of intermediate
resonances is called {\tt Variant-1.0}.
If supplemented with higher-level expert features like
invariant masses of intermediate resonances or energy fractions, it is called
{\tt Variant-1.1.}
  For all three channels results for {\tt Variant-1.0} and
{\tt Variant-1.1} are close. Expert variables provide redundant information only.
In Fig.~\ref{Fig:ML_Variant-All-1.0} (left plot)  AUC results for training and
validation of $ a_1^{\pm} - \rho^{\mp} $ are shown for {\tt Variant-1.0}. The highest result on the validation sample is around $0.60$.

In Fig.~\ref{Fig:ML_Variant-All-1.0}, right plot, we show  ROC curves
 displaying True Positive Rates (TPR) versus
False Positive Rates (FPR) for {\tt Variant-All} and {\tt Variant-1.0}. 

The achieved AUC's and APS's are collected in the respective lines of Table~\ref{Tab:ML_AUC}.
The large gap of AUC and APS performance between   {\tt Variant-All} and  {\tt Variant-1.0} feature sets, is present for all
channels. In the following, we attempt to improve performance thanks to 
 information on the neutrino momenta and in particular their azimuthal angles.

\begin{figure} 
  \begin{center}                              
    { 
      \includegraphics[width=7.5cm,angle=0]{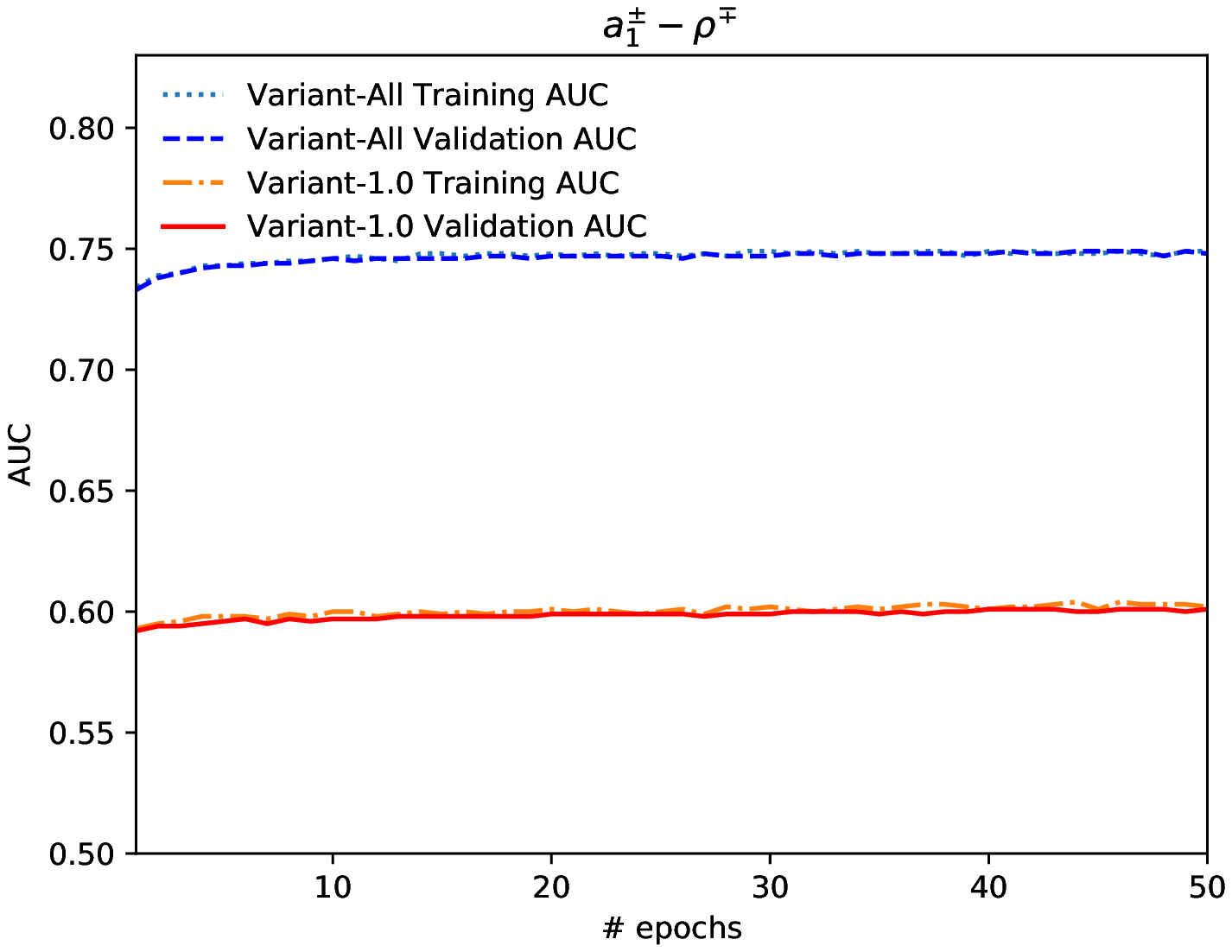} 
      \includegraphics[width=7.5cm,angle=0]{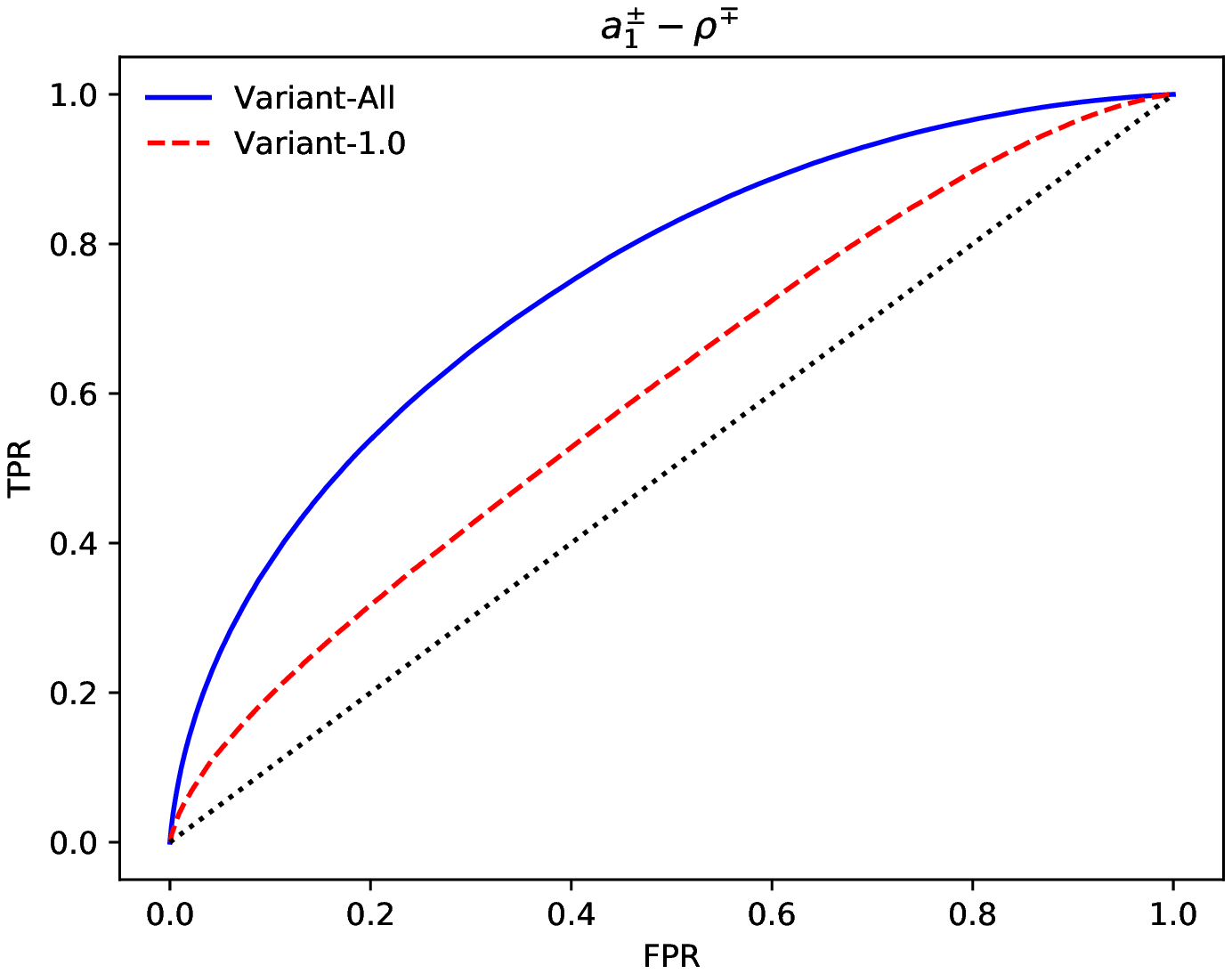}
    }
  \end{center}
  \caption  {
    The AUC score (left plot)  for training and validation of $a_1^{\pm}-\rho^{\mp}$ samples,
    as a function of number of epochs and ROC curve for 50 epochs (right plot). The 
    {\tt Variant-All} and {\tt Variant-1.0} 
    were chosen for the list of features. Training and validation curves overlap.
    \label{Fig:ML_Variant-All-1.0}
    }
\end{figure}

\subsection{Adding neutrino momenta}

In this Subsection we present improvements due to the energy and longitudinal
neutrino momenta. Such an extension of the features list is not expected
to be very beneficial, as CP information is carried by the transverse degrees
of freedom, but it may optimize the use of 
information learned from correlations of hadronic decay products.

With assumptions explained in Section~\ref{Sec:addNeutrina}, we  approximate
each of neutrino momentum components $E_{\nu}$, $p^z_{\nu}$,
$p^T_{\nu}$ in the rest-frame of hadronic decay products. It is interesting
to check first
what is the potential impact of that information, i.e.  when
truth level  values are used. 
We 
add the laboratory frame  $E^x_{miss}, E^y_{miss}$,
redundant to some extend, as it was already used in Eq. (\ref{eq:collinear2}) for  $p^z_{\nu}$.

The augmented list of features, using true components of neutrino momenta, is denoted as  {\tt Variant-2.0}, while the ones
using approximate components of neutrino momenta are denoted as {\tt Variant-2.1} and {\tt Variant-2.2}, depending on whether the information on
$E^x_{miss}, E^y_{miss}$ is included or not. The AUC  and APS scores by the {\it DNN} 
for $\rho^{\pm}-\rho^{\mp}$, $a_1^{\mp}-\rho^{\pm}$ and $a_1^{\pm}-a_1^{\mp}$ channels
are displayed in Table~\ref{Tab:ML_AUC}. The improvement from  {\tt Variant-1.0} to
{\tt Variant-2.0} is not impressive. We observe later a small  performance
degradation  from {\tt Variant-2.0} to
{\tt Variant-2.1}, which uses approximate
neutrino features resulting in sensitivity loss. The laboratory frame
$E^x_{miss}, E^y_{miss}$ of  {\tt Variant-2.2} 
are, as expected, of no help. In Fig.~\ref{Fig:ML_approx} we show DNN performance for the  $a_1^{\pm}-\rho^{\mp}$ samples as a function of number of training
epochs: the AUC achieved as a function of number of epochs  and the ROC curves.

For the feature sets: {\tt Variant-2.1} and {\tt Variant-2.2}, all
three different approximations for
  $E_{\nu}$, $p^z_{\nu}$, $p^T_{\nu}$  were studied.
The differences between {\it Approx-1}, {\it Approx-2} and {\it Approx-3}
are small but will
certainly show once detector effects are included.

\begin{figure} 
  \begin{center}                              
    {
      \includegraphics[width=7.5cm,angle=0]{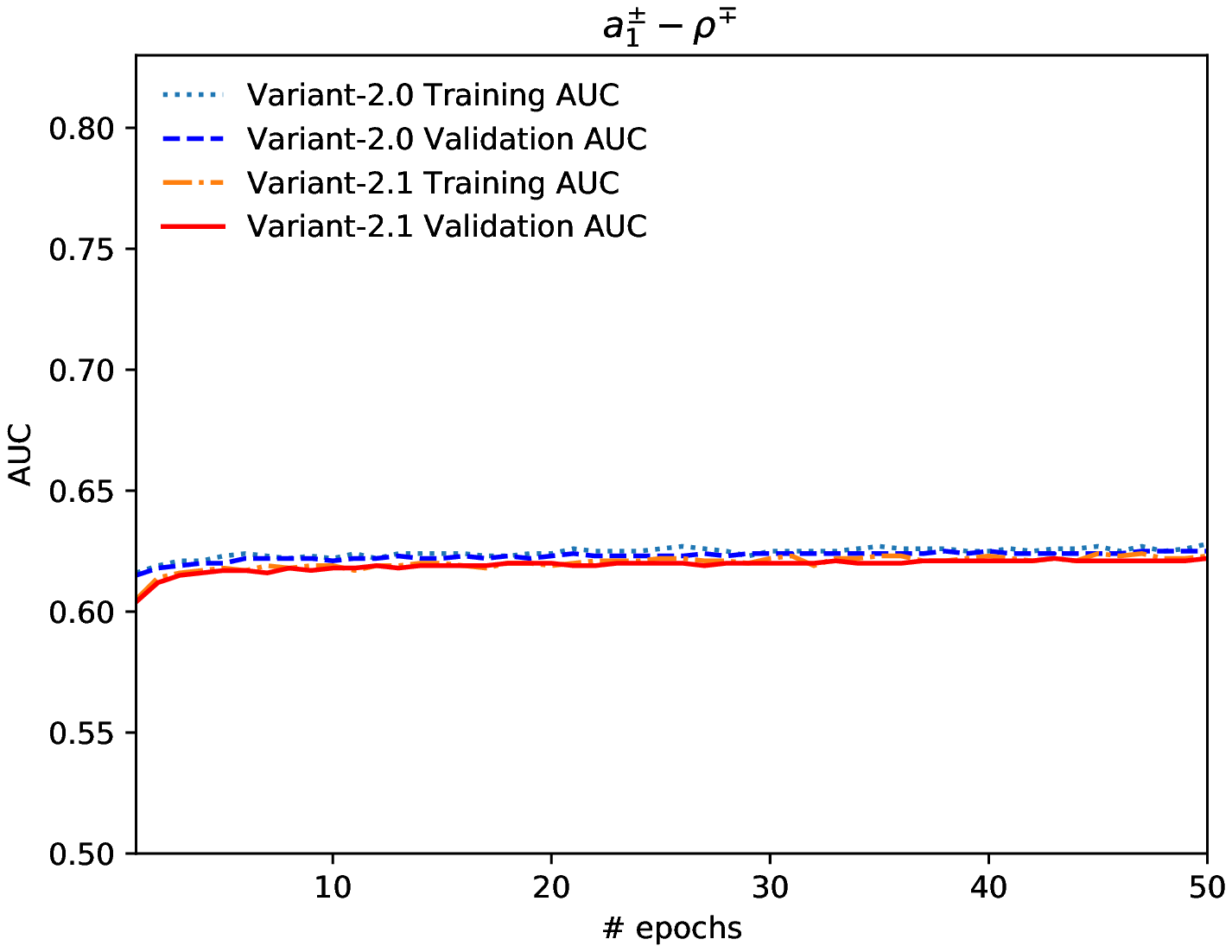}
      \includegraphics[width=7.5cm,angle=0]{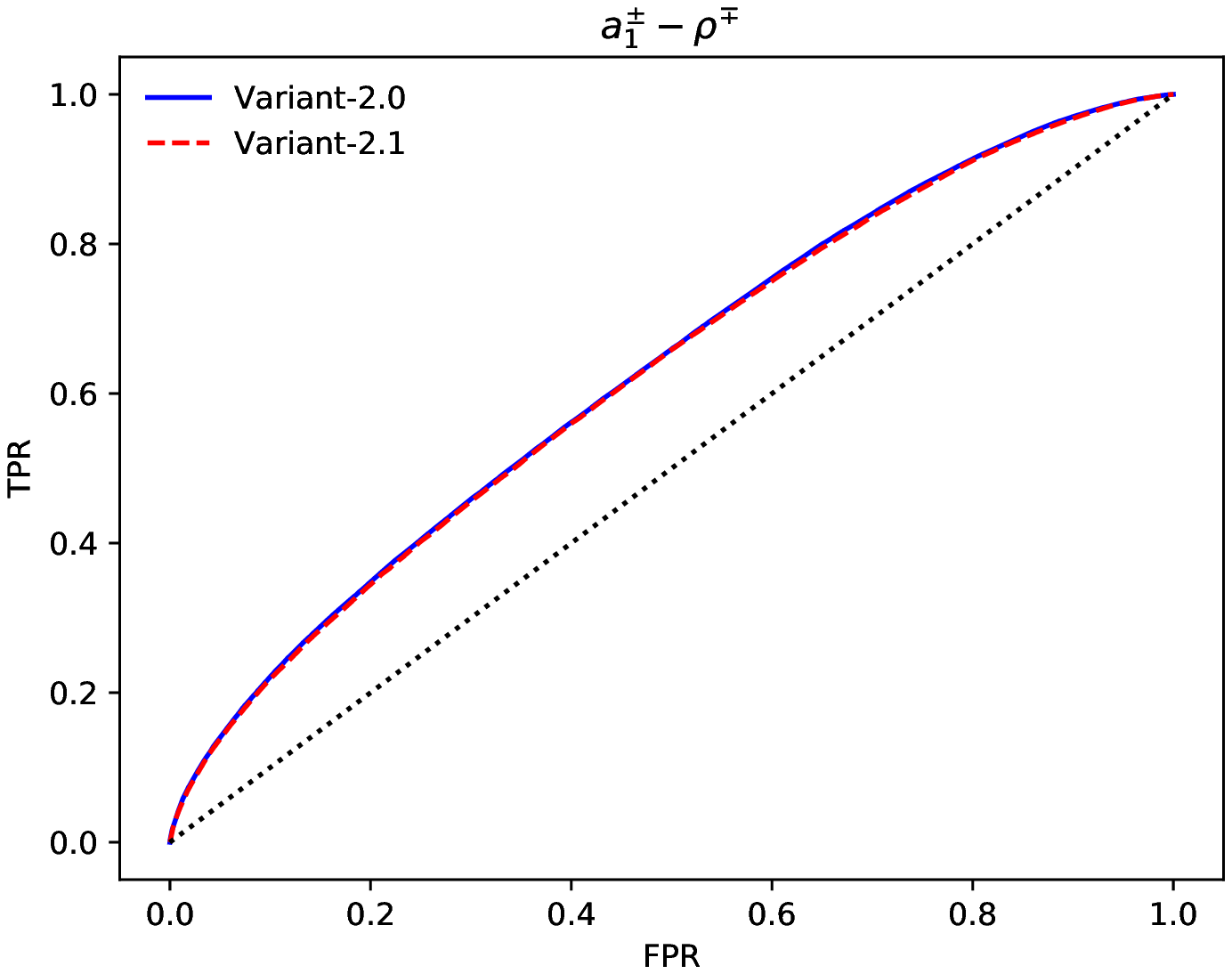} 
    }
  \end{center}
  \caption  {
   Left plot: the AUC score for  training and validation of $a_1^{\pm}-\rho^{\mp}$ samples,
    as a function of number of epochs. The 
    {\tt Variant-2.0}  and {\tt Variant-2.1}  were chosen for the list of features.
    Right plot: the corresponding ROC curves for 50 epochs.
   \label{Fig:ML_approx}
  }
\end{figure}


Clearly, the improvement from approximated information on the neutrinos energy and momenta
(longitudinal and module of transverse)
 is rather small for all three channels.
The most sensitive information on the CP state lies in
 azimuthal angles
of the individual neutrinos. That is in individual
$p^x_{\nu}, p^y_{\nu}$ components of 
hadronic decay products rest-frame
and not in  $p^T_{\nu} = \sqrt{ (p^x_{\nu})^2+  (p^y_{\nu})^2 }$.
Realistically any information on the individual $p^x_{\nu}, p^y_{\nu}$
could be reconstructed
only if the measurement of the $\tau$ decay vertices was possible. In the next Section, we  evaluate how accurately
this information has to be known to become useful.
It constitutes a separate experimental challenge.
Note that at this step,  all components of $\nu_{\tau}$ momenta
except individual $p^x_{\nu}, p^y_{\nu}$ are  reconstructed
sufficiently well
from the measurable quantities.

\subsection{Azimuthal angles of neutrinos from decay vertices}
\label{Sec:addIP}

The azimuthal angles $\phi_{\nu_1}$, $\phi_{\nu_2}$ can be obtained from the measurement of the
$\tau$ lepton decay vertices.  It allows to
reconstruct  the $\tau$-lepton momenta and hopefully can be used for our purpose as well.
This is rather widely used technique in the experimental measurements,
see e.g.~\cite{Jeans:2015vaa},
but so far for $\tau$-mass or $\tau$-lifetime measurement rather than for neutrino azimuthal angles.

We do not aim to reconstruct those angles, we simply calculate them from the
neutrinos 4-momenta and add to the feature lists~\footnote{
  The sub-sub-index $\beta$ encodes the size of the smearing parameter $\beta$.}
 {\tt Variant-3.0} and  {\tt Variant-3.1.$\beta$}. The first one is when the true
 $\phi_{\nu_1}^{true}$, $\phi_{\nu_2}^{true}$ are used, and the second one is
 with smeared  $\phi_{\nu_1}^{smear}$, $\phi_{\nu_2}^{smear}$. In
 Fig.~\ref{Fig:DeltaPhiSmear} the
 $\phi_{\nu}^{true}-\phi_{\nu}^{smear}$ distribution for $\beta=0.4$
 of Eq.~(\ref{eq:IP}) is shown.

The AUC scores are evaluated for the  $\beta$  in  $(0, 2)$
range. In Fig.~\ref{Fig:ML_IP_summary} the  AUC's  for  test samples
of the three channels
   are given  as a function of 
$\beta$. 
The AUC scores for  $\beta=0.0$ reproduce, as they should, the ones of {\tt Variant-3.0} and
are not very far from  scores of  {\tt Variant-All}. That is because
the only difference is  approximate information
on energy,  longitudinal and  transverse momenta of the neutrino.
For  $\beta$ above 1.4,
the AUC  decrease to the ones of {\tt Variant-2.1} sets, which is then
equivalent to not having information
on the neutrino azimuthal angles at all. Even
$\phi_{\nu_1}^{smear}$, $\phi_{\nu_2}^{smear}$, corresponding to
rather large $\beta=0.$4
contributes sizably to CP Higgs sensitivity.
 \begin{figure}
 \begin{center}                              
 { 
   \includegraphics[width=7.5cm,angle=0]{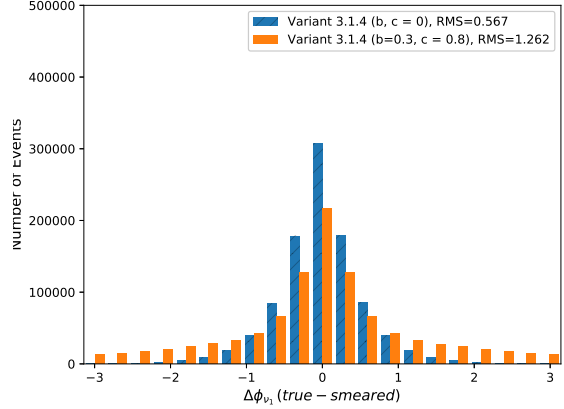}
 }
\end{center}
 \caption  { The (true - smeared) distribution of $\Delta \phi_{\nu_1}$  for  $\beta$=0.4
    with/without additional polynomial modulation of Eqs.~(\ref{eq:IP_abc})/(\ref{eq:IP}).
 }\label{Fig:DeltaPhiSmear}
\end{figure} 
\begin{figure} 
  \begin{center}                              
    { 
      \includegraphics[width=7.5cm,angle=0]{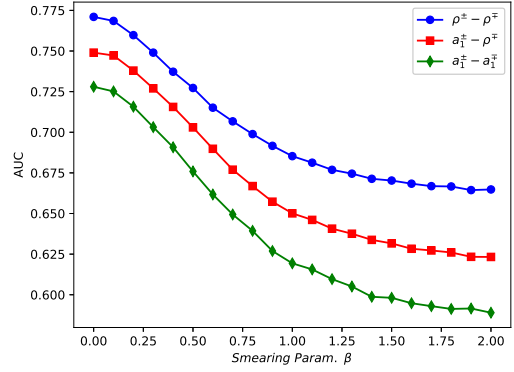}
    }
  \end{center} 
  \caption {The AUC in function of smearing parameter $\beta$ of Eq.~(\ref{eq:IP_abc})
    for $\rho^{\pm}-\rho^{\mp}$, $a_1^{\pm}-\rho^{\mp}$ and $a_1^{\pm}-a_1^{\mp}$ channels
    and $Variant-3.1.\beta$.
    \label{Fig:ML_IP_summary}}
\end{figure}		
The derivative of the sensitivity with respect to $\beta$, reaches its maximum  at about $0.35$ and remains constant
until  $\beta=0.9$. Then nearly all sensitivity gain is lost.  
For even larger $\beta$, loss of sensitivity continues, but as the contribution
is then already small,
deterioration is small too.


Let us now  check if the {\it DNN} algorithm is sensitive to precise
modeling of the $\phi_{\nu_{1,2}}$ resolution.  That is why, for
the validation and test sample, we 
introduce~\footnote{Polynomial  modification is implemented in validation and test samples with the help of Monte Carlo  unweighting:
$ wt = f_{valid}(\Delta \phi_{\nu}, \beta, b, c)/f_{train}(\Delta \phi_{\nu}, \beta)$
 . 
 }
an additional  polynomial component  for the smearing
\begin{equation}
  f_{valid}(\Delta \phi_{\nu},\beta,b,c) =  f_{train}(\Delta \phi_{\nu},\beta) \bigl( 1 + b^2  \Delta \phi_{\nu}^2 +  c^2 \Delta \phi_{\nu}^4 \bigr).\nonumber
  \label{eq:IP_abc}
\end{equation}
The results should mimic the impact of inefficiencies (mismodeling) of the {\it DNN}
training sample with respect to
what is present in the validation or test samples.
In Fig.~\ref{Fig:DeltaPhiSmear} the distribution of  $\phi_{\nu}^{true}-\phi_{\nu}^{smear}$ is given for $\beta=0.4,$ and $b,c=0.3,\;0.8$.
In Table~\ref{Tab:IP_beta} results for $\rho^{\pm}-\rho^{\mp}$, $a_1^{\pm}-\rho^{\mp}$, $a_1^{\pm}-a_1^{\mp}$ channels
and for $\beta$ = 0.2, 0.4, 0.6 with further choices of $ b$ and $c$ are collected. 
The additional polynomial component of smearing introduced to the test sample is not affecting
the {\it DNN} performance.
We can see that the degradation due to  $b,c=0.3,\;0.8$ is small
and the results  provide some encouraging insight to the  {\it DNN} capacity to exploit imprecise information and point to possible direction for the
studies of systematic uncertainities~\footnote{We hope,
  that these degradation parameters
will be replaced in the future by sophisticated detector simulations.
Our evaluation indicates, that already modest and even partial
reconstruction of the $\tau$ decay vertex position is useful.
The experimental effort, like references~\cite{Aad:2015unr, Sirunyan:2018pgf, Aad:2015ydr, Sirunyan:2017ezt} and as mentioned
in~\cite{Cherepanov:2018yqb}, is encouraging.
\\
Our example can not substitute future  work with well understood
detection details.  Nonetheless it hints for 
a possible method of  experimental
ambiguities  evaluation in {\it DNN} applications. 
}.

In our study, when
 precision of experimental inputs  was
expected to be better than that from decay vertices impact parameters,
we have
reconstructed neutrino momenta components from hadronic products and conservation laws.
Only the $\phi_{\nu}$ angles  required this
rather low precision input.
From  Fig.~\ref{Fig:ML_IP_summary} we can expect that  approximate $\phi_{\nu}$ angle
with ambiguity of up to $\pm \frac{\pi}{4}$ may sizably improve sensitivity.

Such conjecture on the size of $\phi_{\nu}$ smearing critical for CP sensitivity
is of interest for any ML application.
For $\beta=1.2$ the shift $\Delta \phi_{\nu}$ was bigger than $\frac{\pi}{4}$
 in sizable fraction of events. Then;
DNN solution does not gain sensitivity  from  $\phi_{\nu}$.
Still,  an approach relying less on $\phi_{\nu}$ measurement,
but on restricting, which events should be dropped from the analysis could be useful.
Possibly, for large smearing, elimination of events with high risk of $\phi_{\nu}$
misreconstruction  may be appropriate as it was attempted in Ref.~\cite{Desch:2003mw}.
A discussion of physics properties simultaneously with those of the ML
algorithms may be of interest again.

\begin{table}
 \vspace{2mm}
  \caption{
    The AUC and APS for  $\rho^{\pm}-\rho^{\mp}$, $a_1^{\pm}-\rho^{\mp}$ and  $a_1^{\pm}-a_1^{\mp}$ channels with features {\tt Variant - 3.1.$\beta$}
   ($\beta$ = 0.2, 0.4 and 0.6  is used for training, validation and test samples).
    For the test sample  polynomial modifications of  smearing function,  Eq.~(\ref{eq:IP_abc}), were introduced. \label{Tab:IP_beta} }
  \begin{center}
    \begin{tabular}{|l|c|c|c|}
     \hline
      \cline{1-4}\multicolumn{1}{|c|}{}&\multicolumn{3}{|c|}{  AUC / APS }\\
      \hline   
      Parameters &  ($\beta=0.2$) &  ($\beta=0.4$)&   ($\beta=0.6$) \\
            \cline{1-4}\multicolumn{1}{|c|}{}&\multicolumn{3}{|c|}{  $\rho^{\pm}-\rho^{\mp}$ }\\
      \hline
      b = 0.0, c = 0.0  &  0.761/0.759   &  0.739/737   &  0.715/0.714 \\
      b = 0.3, c = 0.8  &  0.760/0.758   &  0.739/0.736   &  0.716/0.713 \\
      b = 0.9, c = 0.9  &  0.759/0.756   &  0.738/0.734   &  0.714/0.713 \\
      \cline{1-4}\multicolumn{1}{|c|}{}&\multicolumn{3}{|c|}{ $a_1^{\pm}-\rho^{\mp}$ }\\
      \hline	
      b = 0.0, c = 0.0  &  0.739/0.731   &  0.714/0.706   &  0.687/0.679 \\
      b = 0.3, c = 0.8  &  0.738/0.730   &  0.714/0.705   &  0.687/0.679 \\
      b = 0.9, c = 0.9  &  0.737/0.728   &  0.714/0.704   &  0.687/0.678 \\
      \cline{1-4}\multicolumn{1}{|c|}{}&\multicolumn{3}{|c|}{ $a_1^{\pm}-a_1^{\mp}$ }\\
      \hline
      b = 0.0, c = 0.0  &  0.713/0.705   &  0.690/680   &  0.660/0.653 \\
      b = 0.3, c = 0.8  &  0.715/0.706   &  0.693/0.682   &  0.661/0.653 \\
      b = 0.9, c = 0.9  &  0.714/0.706   &  0.688/0.680   &  0.660/0.653 \\      
     \hline
    \end{tabular}
  \end{center}
\end{table}

\subsection{Tau lepton direction}
\label{Sec:addTaudir}

The approximate information on the $\tau$ lepton direction enables the {\it DNN} to
constrain the
neutrino  and  significantly improve the classification.
For that {\tt Variant-4.0} and {\tt Variant-4.1} are defined in
Table~\ref{Tab:ML_variants}.

In  Table~\ref{Tab:ML_AUC}, performances of the {\it DNN} are presented,
when
true level, or from approximation, $\tau$ lepton spatial momenta components in the respective
$\rho^{\pm}-\rho^{\mp}$, $a_1^{\pm}-\rho^{\mp}$ and $a_1^{\pm}-a_1^{\mp}$ rest-frame are added.
We observe significant
improvement of the performance with
respect to {\tt Variant-2.1}  and comparable to {\tt Variant-3.1.X} family.
In fact, the  performance of {\tt Variant-4.0} is close to {\tt Variant-All}. 
{\tt Variant-4.1} is a bit lower, close to {\tt Variant-3.1.4}.
Then only
$\tau$ direction in the
laboratory frame  is exact and the energy
is obtained from the  simple ansatz of Subsection~\ref{Sec:taudir}.
When such $\tau$ 4-momentum is boosted to
$\rho^{\pm}-\rho^{\mp}$, $a_1^{\pm}-\rho^{\mp}$ and $a_1^{\pm}-a_1^{\mp}$ rest-frame,
its direction
absorbs some biases. Results of {\tt Variant-4.1}
indicate that {\it DNN} efficiently converts such an input
into information on $\nu_\tau$.

\section{Summary}
\label{Sec:Summary}

From the perspective of theoretical modeling,
the CP parity phenomenology in cascade $H \to \tau \tau$, $\tau^\pm \to had^{\pm} \nu_\tau $ decay
is  rather simple, because the matrix element can be
easily defined.
On the other hand, the parity effect manifests itself in rather complicated features
of multi-dimensional distributions where kinematic constraints related to
ultra-relativistic boosts  and detection ambiguities play an important role in
the reconstruction of the $\tau$ decay kinematic.
Our aim was to evaluate what level of precision and for each experimentally
available features need to be achieved by experiments for the meaningful
measurements.

In our previous paper~\cite{Jozefowicz:2016kvz} we have studied the performance
of the  DNN binary classification technique
for the  hadronic $\tau$ leptons decay products only. 
Now we have turned our attention also to the $\nu_\tau$ momenta.

Whenever possible, we  have exploited  constraints of $\tau$-mass, $H$-mass and energy momentum conservation to minimize dependence on
highly smeared  neutrino kinematic deduced from the impact parameter of $\tau$ decay and  production vertices.
The resulting set of expert variables helps DNN algorithms to identify physics sensitive
variables useful to identify differences between the  event classes.

Reconstructed with 
approximation but from visible decay products, longitudinal components of the neutrino momenta
alone improved the
AUC from 0.656, 0.609, 0.580 to about 0.664, 0.622, 0.591 respectively for
$\rho^{\pm}-\rho^{\pm}$, $a_1^{\pm}-\rho^{\mp}$ and $a_1^{\pm}-a_1^{\mp}$ cases.
The improvement for the Higgs boson CP sensitivity
is rather minuscule, even when the detector effects were not taken into account.

A more significant improvement came when the transverse components of the neutrino momenta were known,
even imprecisely. This can be achieved if  the $\tau$-lepton
decay vertices are measured and 
used to reconstruct directions of the $\tau$ leptons momenta.
The performance of such reconstruction is detector specific and is a challenge.
We have  estimated how big of an improvement of CP sensitivity
is obtained as a function of detection smearing for the azimuthal angles $\phi_\nu$ and $\phi_{\bar\nu}$. 
Even with  large smearing, $\beta=0.4$, the AUC improved from 0.664, 0.622 and 0.591 to about 0.738, 0.714 and 0.687
for $\rho^{\pm}-\rho^{\pm}$,  $a_1^{\pm}-\rho^{\mp}$ and $a_1^{\pm}-a_1^{\mp}$ cases, respectively.
 Note that   $\phi_\nu$ and $\phi_{\bar\nu}$ angles
 represent an intermediate step in the quest: from expert variables to DNN algorithms with direct
 use of low-level features.
 We are leaving the topic of the angles measurements and use for forthcoming works.

 Similar performance is expected when good quality  $\tau$ lepton
 laboratory frame  direction, as seen in the rest-frame of
 all visible Higgs decay products combined is  available for the evaluation
 of $\tau$ direction. 
 The ambiguity on the laboratory frame $\tau$ energy is not that important.
The enhancement with $\tau^\pm$ directions was achieved ({\tt Variant-4.1}), the AUC reached 
0.738, 0.704 and 0.683 respectively for $\rho^{\pm}-\rho^{\pm}$,   $a_1^{\pm}-\rho^{\mp}$ and $a_1^{\pm}-a_1^{\mp}$
cases.

In Fig.~\ref{Fig:ROC_manyVariants} we show ROC curves for different variants of feauture lists discussed in this paper.

 \begin{figure}
 \begin{center}                              
 { 
   \includegraphics[width=7.5cm,angle=0]{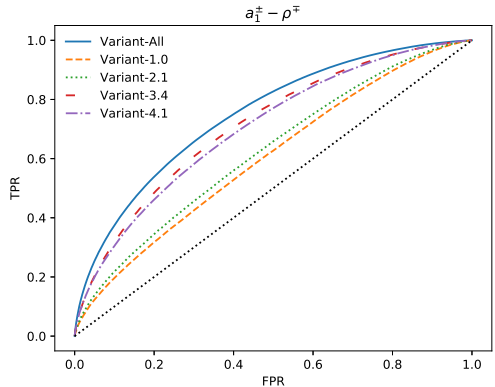}
 }
\end{center}
 \caption  { The ROC curves for different feauture lists.
 }\label{Fig:ROC_manyVariants}
\end{figure} 

 The concept of the {\it optimal observables} is used since many years to obtain 
 phenomenologically sound results.  For ML classification,
 where 
  multi-dimensional input is used, it provides essential tests.
 An approach, where sophisticated methods are used to measure $h_\pm$ of Eq.~(\ref{eq:matrix}),
 should be mentioned. All complexity of hadronic $\tau$ decays
 and detector response is then hidden  in  each $\tau^\pm$ polarimetric vector
 $h_\pm$.
 Once an algorithm for $h_\pm$ reconstruction prepared, the later  step of CP
 phenomenology is straightforward: details of $\tau^\pm$ decay channels and
 detector effects are resolved.
 The $h_\pm$ complexity is smaller than of the whole $H \to \tau\tau$ cascade decay.
 It is independent from the Higgs phenomenology and  preparation 
 can rely on the much more abundant $Z \to \tau\tau$ data. Such a possibility was mentioned in~\cite{Cherepanov:2018yqb}
 and is pursued e.g. by the CMS collaboration.
Then, the ML learning techniques could be used to reconstruct $h_\pm$ vectors
from the complex detector responses to particular $\tau$ decay channels
and details of its decay vertex position.

 The evaluation which of the  methods is best, or in fact,
 how complementary the methods can be, requires work of experimental groups.

 Recently, in Ref.~\cite{Erdmann:2018shi}, classifiers specifically tuned to tackle
 the Lorentz group features of High Energy Physics signatures were
 prepared and used. This could be useful for {\tt Variant-1.0}, where
 four momenta of secondary $H\to \tau\tau$  decay products are used only. In the present work
 this may be less straightforward as part of the features is intimately related
 to laboratory frame  and their transformation to other frames
 may be poorly defined. That is why, expert variable style
 reconstruction of
 neutrinos azimuthal angles may be an efficient way to follow, or at least useful
 to better understand limitations and ambiguities of  methods.

 \vskip 0.3 cm
\centerline{\bf \large Acknowledgments}
\vskip 0.2 cm
P. Winkowska would like to thank L. Grzanka for valuable comments and suggestions through the
time of preparing results.

This project was supported in part from funds of Polish National Science
Centre under decisions DEC-2017/27/B/ST2/01391.
DT and PW were supported from funds of Polish National Science Centre under
decisions DEC-2014/15/B/ST2/00049.

Majority of the numerical calculations were performed at the PLGrid Infrastructure of 
the Academic Computer Centre CYFRONET AGH in Krakow, Poland.


\appendix

\section{Deep  Neural Network }
\label{App:DNN}

The structure of the simulated data and the {\it DNN} architecture follows what was published in our previous paper~\cite{Jozefowicz:2016kvz}.
It is prepared for   {\tt TensorFlow}~\cite{abadi2015tensorflow}, an open-source machine learning library. The
learning procedure is optimized using a variant of the stochastic gradient descent algorithm called Adam~\cite{kingma2014adam}.
We also use  Batch Normalization~\cite{ioffe2015batch}
(which has regularization properties) and Dropout~\cite{srivastava2014dropout}
(which prevents overfitting) 
to improve the training of the {\it DNN}. The problem of determining Higgs boson CP state is framed as binary classification
because the aim is to distinguish between the two possible scalar and pseudoscalar Higgs CP states.

%
We consider three separate problems for $H\to \tau \tau$  channels:
$\rho^{\pm}-\rho^{\mp}$, $a_1^{\pm}-\rho^{\mp}$, $a_1^{\pm}-a_1^{\mp}$. 
We solve all three problems using the same neural network architecture.
Depending on the decay channel
for the outgoing $\tau$ pairs, each of the cases contains different number
of dimensions to describe an event, i.e. production of the Higgs boson decaying 
into $\tau$ lepton pair.
Each data point consists of features which represent the observables/variables
of the consecutive event.
The data point is thus an event of the Higgs boson production and decay into $\tau$ lepton pair. 
The structure of the event is represented as follows:
\begin{equation}
  x_i = (f_{i,1},...,f_{i,D}),w_{a_i},w_{b_i}
\end{equation}
The  $f_{i,1},...,f_{i,D}$ represent numerical features  and $w_{a_i},w_{b_i}$ are weights proportional to the likelihoods
that an event comes from a set $A$ or $B$ (binary scalar or pseudoscalar classification). The weights  calculated
from the quantum field theory matrix elements are available and stored in the simulated data files.
This is a convenient situation, which does not happen in many other cases of ML classification.
The $A$ and $B$ distributions highly overlap in the $(f_{i,1},...,f_{i,D})$
space, a more detailed discussion
can be found in~\cite{Jozefowicz:2016kvz}.  
The perfect separation is therefore not possible and $w_{a_i}/(w_{a_i}+w_{b_i})$ corresponds to the Bayes optimal probability
that an event is sampled from set $A$ and not $B$. The $w_{a_i},w_{b_i}$ are used to compute targets during the
training procedure.

Because model A and B samples are prepared with the same events and differ with 
the spin weights $w_{a_i}, w_{b_i}$ only, the statistical fluctuations of the learning procedure are largely reduced.
It has also consequences for the actual implementation of the DNN metric and code.

To quantify classification performance,  weighted AUC  
and APS
were used~\footnote{
  In Ref.~\cite{Bialas:2018nvm} we have demonstrated that {\it for our applications} the AUC score approximate reasonably well the probability to identify
  single event as scalar or pseudoscalar, thus  can be used to calculate
  statistical significance of the event sample as well.
},  we have not followed further alternatives.
For each data point $x_i$, the {\it DNN} returns probability $p_i$ that it is correctly (not correctly) classified as of type A
and it contributes to the final loss function twice, with weight $w_{a_i}$ and $w_{b_i}$ respectively. With this definition,
the AUC = 0.5 would be obtained for random assignment, while the AUC = 1.0 would be reached for perfect separation.
As in the studied problems distributions are overlapping, the best achievable AUC$\simeq 0.78$  is reached only with 
 $ p_i = w_{a_i}/(w_{a_i}+w_{b_i}) $ (oracle predictions).
The value depends slightly on the case studied, due to applied minimal set of cuts on the kinematics of $\tau$ decay products
to partly emulate detector conditions.

Weighted events with  $w_a,w_b$ are used for implementation convenience
and to limit statistical fluctuations.
We have repeated some of the {\it DNN} classification chain (training, validation, testing)
using unweighted events~%
\footnote{Randomly sampling events with
 weights  for model A and B. Then  only  such events were used for training.} too.
%
We have found very good consistency of  performance achieved in this way.

The {\it DNN} architecture~\footnote{The choice of the {\it DNN} architecture was optimised with not presented here  studies. A variety
  of the activation functions, number of layers and number of nodes was tried.
The configuration used in~\cite{Jozefowicz:2016kvz} was confirmed  as optimal one.},
 consists of
D-dimensional input (list of features) followed by six layers of 300 nodes each with
ReLU \cite{DBLP:journals/corr/abs-1803-08375} activation functions and 1-dimensional output layer
returning probability $p_i$ of the choice. It is calculated using the softmax function.
The metric minimized by the model is negative log likelihood of the true targets under Bernoulli distributions.

The parameter which was optimized with respect to what was used in~\cite{Jozefowicz:2016kvz}
was a dropout~\cite{srivastava2014dropout}. For the analysis presented in Ref.~\cite{Jozefowicz:2016kvz}
 AUC  score  was obtained after
 training with 5 epochs. It was considered sufficient. Here, given
 the variety of feature lists, we performed training
on much larger number of epochs and studied what would be the optimal working point for number of epochs and dropout.
The 20\% dropout seemed to be the optimal choice to avoid overfitting,
which occurs in the case of a larger number of training epochs.
The best performance was achieved after 5-50 epochs, depending on the case.
Training with 50 epochs was used on the test samples to quote the 
final AUC score.  While optimizing dropout level, we have observed that although it leads to more
robust  {\it DNN} (smaller risk for overfitting), the performance was sometimes somewhat reduced. The positive impact
of dropout is illustrated (on dataset consiting of 1M event samples)  
in Fig.~\ref{FigApp:DNN_dropout}  for  $\rho^{\pm}-\rho^{\mp}$ channel and {\tt Variant-1.1}
of training and validation.

 \begin{figure} 
  \begin{center}                              
    { 
      \includegraphics[width=7.5cm,angle=0]{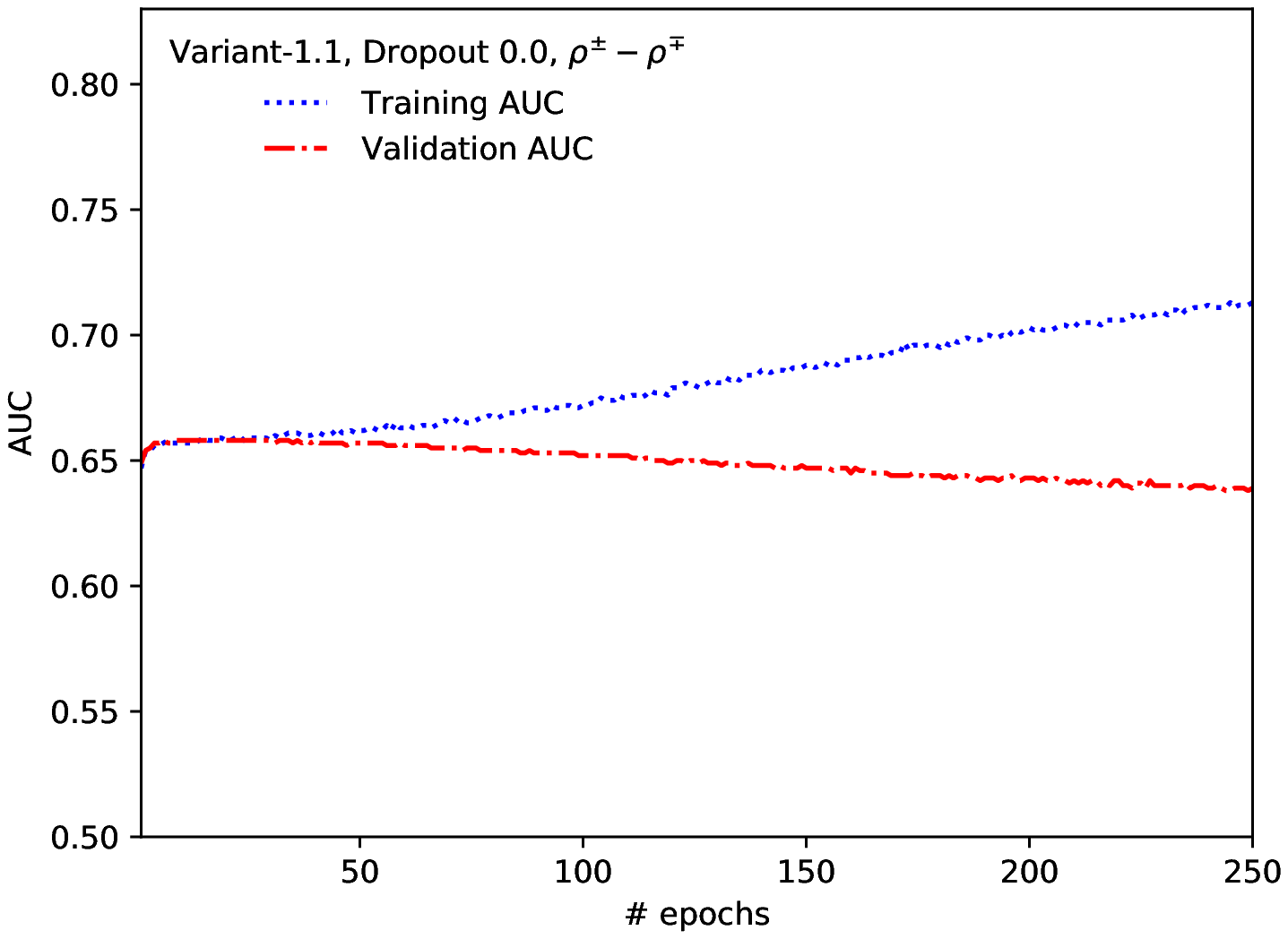} 
      \includegraphics[width=7.5cm,angle=0]{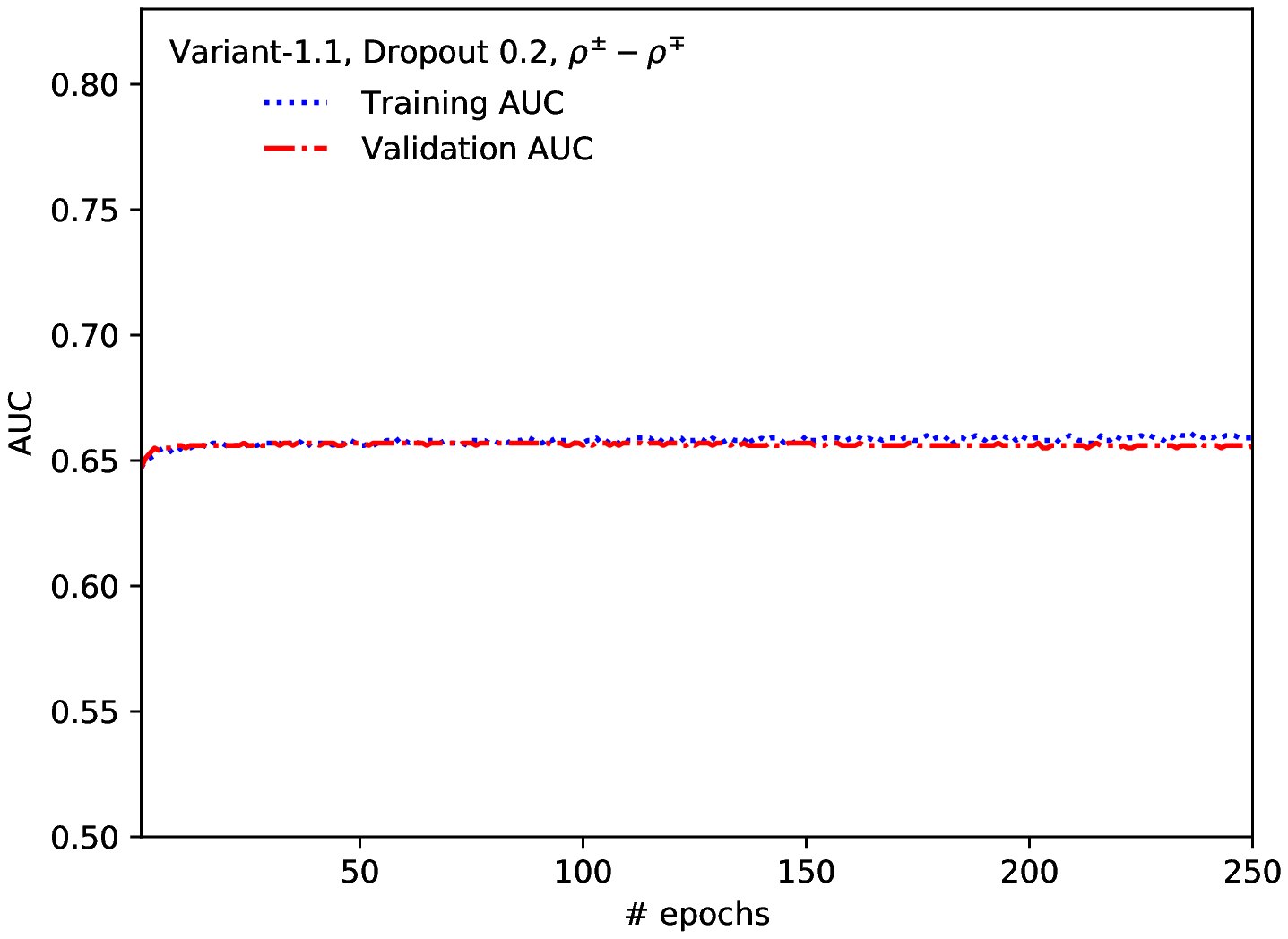}
    }
  \end{center}
  \caption  {
    The  {\it DNN} training and validation  AUC
    as a function of number of epochs, for $\rho^{\pm}-\rho^{\pm}$ channel and  sample features {\tt Variant-1.1},
    without dropout (left plot) and with dropout $= 0.2$ (right plot).
    \label{FigApp:DNN_dropout}
    }
\end{figure}

\section{Alternative ML techniques}
\label{App:MLdivers}

Although Deep  Neural Networks are often used for classification tasks in
High Energy Physics, many other more classical techniques are used as well and
often are able to achieve  similar classification performance. Despite promising results that are
often enlisted in papers, one should always remember that a Machine Learning
technique that  perform well on one data-set with specific features,
may deliver not so promising results on the other.

This observation  is of  fundamental
nature and results from the  mathematical assumption behind particular ML libraries.
The solutions which were developed for the libraries,
depend on the application domains the particular systems were prepared for. 
Ref.~\cite{Delgado14a} can be used as a guidance. The
recent study~\cite{comparison} collected extensive comparison of several  Machine Learning algorithms.
Despite not being able to
contribute much to that topic,  we  show
that for the application discussed through this paper it is  indeed the case:
the {\it DNN} technique  by far
outperforms  the more classical approaches.


The following   Machine Learning techniques were chosen for the comparative study:
\begin{itemize}
\item
  {\it  Boosted Trees} {\it (BT)}~\cite{Chen:2016btl}
\item
  {\it Random Forest} {\it (RF)}~\cite{Breiman:2001hzm}
\item
  {\it   Support Vector Machine} {\it (SVM)}~\cite{Bevan:2017stx}
\end{itemize}
For {\it Boosted Trees} the  {\tt XGBoost}~\cite{Chen:2016btl} library was used
while for {\it SVM} and {\it Random~Forest} the  {\tt scikit-learn}~\cite{ScikitLearn2011}
was chosen.

The  AUC score was used to evaluate performance of ML methods.
This is one of the recommended approaches when using a single
number in evaluation of Machine Learning algorithms on binary classification problems. 
To  minimize bias
the comparisons were
carried out on the same data-sets~\footnote{In case of {\it SVM} subsets were used due to CPU limitations.}.

For the  {\it Boosted~Trees} method the point of interest was to check the dependence of obtained results on the depth of a tree.
The AUC score as a function 
of a tree depth is given on Fig.~\ref{Fig:boostedTreesDepth}  for the  $\rho^{\pm}-\rho^{\mp}$ case
 and several variants of the feature list.
As bigger depth affects complexity of computation,
a search for optimal choice  both in terms of  results and efficiency of computation
was performed. Tree depth between 3 and 10 is suggested in~\cite{Chen:2016btl}.
At
first  we have used depths from 3 to 20. The upper bound was increased 
 to see the trend on  AUC score plots.
The results seemed to rise up to 20.  Additional evaluation with the  depth equal
to number of features of a given {\tt Variant-X.Y} was also carried out.

As suggested in the literature \cite{howmanytreesinrandomforest},
for the  {\it Random~Forest} method
128 trees (estimators)  were used.
For the 
next best split during the tree building number of features 
equal to $log_2(N_{f})$ or $\sqrt{N_f}$, where $N_{f}$ denotes number of features, was tried.
Performances of the two choices were comparable. For the trees depth the
optimal value, as found  for  {\it Boosted~Trees} was used.
Tests with a larger number of trees (300) and bigger tree depths (30)
  were also carried out.

  For the  {\it Support Vector Machine} method, first tests were performed
  to determine which kernel,
linear or RBF,  gives more promising results.
This resulted in choice of RBF kernel with better results  stability.
In the next step fine-tuning
of $C$ and $\gamma$ parameters (soft margin and kernel parameter) was performed.
The parameters evenly distributed on logarithmic scale from $10^{-3}$ to $10^{3}$
were tested. To avoid excessive computation, fine-tuning was performed
only for  {\tt Variant-All},
and on smaller event sample. The obtained parameters, $C$ = 10 and
$\gamma$ = 0.1,
  were then used
to train the classifier on other feature lists as well.  

The comparison of best performance for the  $\rho^{\pm}-\rho^{\mp}$ channel and different variants
of  features is shown in Table~\ref{AppTab:ML_other}.
Clearly performance of the {\it DNN} is outstanding.  

\begin{table}
\vspace{2mm}
\caption{AUC score for different ML methods and different feature sets
  obtained on  validation $\rho^{\pm}-\rho^{\mp}$ sample.}
\label{AppTab:ML_other}
\begin{center}
\begin{tabular}{|l|c|c|c|c|}
\hline
Features set     &  {\it DNN}  &  {\it BT} &  {\it RF} & {\it SVM} \\ \hline
{\tt Variant-All} & 0.764  & 0.641  & 0.626 & 0.635 \\
{\tt Variant-1.0} & 0.655  & 0.569  & 0.567 & 0.559 \\
{\tt Variant-2.1} & 0.657  & 0.564  & 0.564 & 0.572 \\
{\tt Variant-2.2} & 0.657  & 0.562  & 0.559 & 0.568 \\ \hline
\end{tabular}
\end{center}
\end{table}

We have compared execution times, memory usage and efficiency
of all the classifiers.
We have used  the  {\tt Prometheus} cluster~\cite{PrometheusCYFONET}.
Jobs were executed  at 1 node with 4 tasks per node and
5GB memory per task. The comparison is reported in Table~\ref{AppTab:ML_other_CPU}. 
The {\it SVM} training took
by far the longest time. Training of {\it BT} took the least time, under 10 minutes, which made it
 8 times faster to train than  {\it DNN}. Both {\it DNN} and {\it RF} 
used the resources well, achieving
efficiency of 68.8\% and 96.1\% respectively.

Training time  largely vary between the ML methods.
 This is probably due to unexpected by ML relations
 between features.
 The Higgs CP state classification turned out to be a challenge
 for more classical ML algorithms.


\begin{table}
  \caption{Time of training, memory usage and efficiency information
     as reported by job submission  system.
     The $\rho^{\pm}-\rho^{\mp}$ sample with the {\tt Variant-1.0} features
     was used  and different ML methods  were compared.
    The set of 800k training data points was used for {\it DNN},
   {\it BT} and {\it RF}, while for  {\it SVM} only 100k data-points were used. 
  }
\begin{center}
\label{AppTab:ML_other_CPU}
\begin{tabular}{|l|c|c|c|c|}
\hline
 Method & \# Data& Training time & Memory usage & Efficiency\\
&  points &(h:min:sec) & (\% of 20GB) & (CPU)  \\ \hline
{\it DNN}    & 800 k     & 01:21:58 &  6.3\% & 68.8\% \\ 
{\it SVM}                    & 100 k      &  01:52:03 &  4.3\% & 25.0\% \\ 
{\it BDT}  & 800 k    & 00:09:40 &  9.6\% & 24.9\% \\ 
{\it RF}           & 800 k     & 00:34:25 & 37.9\% & 96.1\% \\ \hline
\end{tabular}
\end{center}
\end{table}

\begin{figure}[!ht]
  \centering
    \includegraphics[width=0.5\textwidth]{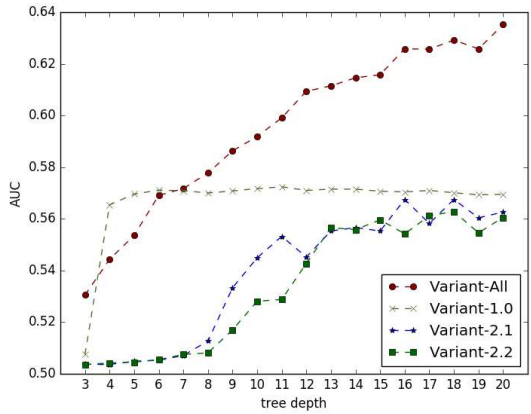}
  \caption{The AUC score of {\it Boosted Trees} classifier as a function  of the tree depth.}
  \label{Fig:boostedTreesDepth}
\end{figure}


\providecommand{\href}[2]{#2}

\end{document}